\documentclass [12pt]{article}         %
\usepackage {apalike,amsfonts}%

\usepackage{cancel} 

\begin{document}

  \author{J. Brian Pitts \\ Faculty of Philosophy and Trinity College \\ University of Cambridge \\ jbp25@cam.ac.uk\\  ORCID 0000-0002-7299-5137 }

\title{General Relativity, Mental Causation,  and Energy Conservation} 
\maketitle 



\abstract{The conservation of energy and momentum have been viewed as undermining Cartesian mental causation since the 1690s.  Modern discussions of the topic tend to use mid-19th century physics, neglecting both locality and Noether's theorem and its converse.  The relevance of General Relativity (GR) has rarely been considered.  But a few authors have proposed that the non-localizability of gravitational energy and consequent lack of physically meaningful local conservation laws answers the conservation objection to mental causation:  conservation already fails in GR, so there is nothing for minds to violate.

This paper is motivated by two ideas.  First, one might take seriously the fact that GR formally has an infinity of rigid symmetries of the action and hence, by Noether's first theorem, an infinity of conserved energies-momenta (thus answering Schr\"{o}dinger's 1918 false-negative objection).   Second, Sean Carroll has asked (rhetorically) how one should modify the Dirac-Maxwell-Einstein equations to describe mental causation.  This paper uses the generalized Bianchi identities to show that General Relativity tends to exclude, not facilitate, such Cartesian mental causation.  In the simplest case, Cartesian mental influence must be spatio-temporally constant, and hence $0$.  The difficulty may diminish for more complicated models. Its persuasiveness is also affected by larger world-view considerations. 

The new general relativistic objection provides some support for realism about gravitational energy-momentum in GR (taking pseudotensor laws seriously).  Such realism also answers an objection to theories of causation involving conserved quantities, because energies-momenta would be conserved even in GR.}



\vspace{.2in}
Keywords: General Relativity, gravitational energy, conservation laws, Noether's theorem, transfer theory of causation, Cartesianism, dualism, interactionism, mind-body problem

\vspace{.1in}


\section{Introduction}

The energy conservation objection to nonphysical mental causation has been made starting from Leibniz in the 1600s \cite{LeibnizNewSystem1695FirstExp} \cite[p. 156]{LeibnizTheodicy} \cite[p. 172]{SchmaltzDescartesCausation} \cite[Book I, chapter 1, section 73]{LeibnizNewEssays} until today and has engaged Wolff, Knutzen, Crusius, the young Kant, Maxwell, Helmholtz, Broad (not all on the same side) and others.  This objection is made with great frequency and often considerable confidence in the contemporary philosophy of mind literature (see citations in \cite{MonteroEnergy,CollinsEnergy,GibbConservation,EnergyMental}).   
Even  E. J. Lowe  thought that conservation laws might be problematic for interactionist dualism, though he proposed several ways out, one of which (though not his favorite)  has merit  \cite{LowePsychophysical} \cite[pp. 56-61]{LoweExperience}  \cite[p. 139]{LoweInvisibility} \cite{EnergyMentalCucuLowe}.
As various authors have noted, non-epiphenomenalist forms of property dualism suffer from analogous worries about how the mental can affect the physical \cite[pp. 40, 43, 50]{CraneElements} \cite[pp. 44-46]{SearleMind} \cite{ZimmermanDualism}.  
Thus one needn't be attracted to substance dualism to encounter this worry.  Unfortunately quite a few dualists, not grasping the relevance of the converse of Noether's first theorem or of the locality of conservation laws in modern physics \cite{EnergyMental}, have tried to argue that mind-to-body causation is compatible with the conservation laws.  Non-conservation is a fact about interactionist dualism, at least insofar as physics is described by the principle of least action (which ignores quantum mechanics).  But to what extent is such non-conservation an objection rather than a just a consequence that can be accepted?

In a previous paper and its companion \cite{EnergyMental,EnergyMentalCucuLowe}, this 300+-year-old objection was assessed in light of modern physics, including aspects that rarely come up and some that seem not to have entered the debate at all.  It was found that, construed as an argument, the objection  begs the question.  Such a conclusion is not entirely novel \cite{WatkinsPhysicalInfluxCrusius,DucasseDualism,AverillKeating,GarberLeibnizEnergy,LowePsychophysical,PlantingaMaterialism,EnergyMentalHistory}, but evidently it  has not been argued in such detail and with an eye to the 20th century physics of fields. On occasion it has been noted that theoretical physics infers conservation laws from symmetries \cite{PlantingaMaterialism}.  This is the topic of Noether's first theorem and its simpler antecedents.  In contrast to Einstein \cite{GorelikConservation,EinsteinEnergyStability}, Max Born, generalizing ideas from Gustav Mie's electrodynamics and Gustav Herglotz's relativistic continuum mechanics, had a clear understanding of how rigid translation symmetries imply conservation laws: 
\begin{quote} 
The  assumption of Mie just emphasized, that the function $\Phi$ [the Lagrangian density]  is independent of
$x,$ $y,$ $z,$ $t,$ is also the real mathematical reason for the validity of the
momentum-energy-law.
\ldots  We assert that for these
differential equations, a law, analogous to the energy law (3$^{\prime}$) of Lagrangian
mechanics, is always valid as soon as one of the 4 coordinates $x_{\alpha}$ does not appear
explicitly in $\Phi$.   \cite{BornEnergyMieHerglotz} \end{quote} 
This result had 19th century antecedents    \cite[pp. 233, 234]{LagrangeEnglish}  \cite[p. 318]{LagrangeMecaniqueAnalytiqueNouvelle}
  \cite{HamiltonConservation,Jacobi,KastrupNoetherKleinLie}.
Clearly interactionist dualist mental causation  violates the symmetries of time- and space-translation.  My soul (if I have one) acts in my body today, not 500 years ago or on the Moon; your soul (if you have one) is analogous.  Thus interactionist mental causation  removes the theoretical expectation of conservation by falsifying the symmetry that would otherwise hold.  The preceding paper novelly applied the \emph{converse} Noether theorem, which says that conservation laws imply symmetries, to show that one can go further \cite{EnergyMental}.  Not only is there no reason to expect conservation to hold, but there \emph{is reason to expect it not to hold}, given interactionism.  Hence given the dialectic of the attempted  objection to interactionism, the critic needs to give some reason for expecting conservation to hold that \emph{has some purchase on the interactionist}.  But there is apparently none to be had, at least not from physics (as opposed to empirical neuroscience), or at any rate not from classical field theories excluding General Relativity.  Hence the conservation argument against interactionism degenerates into a mere denial or an incredulous stare.  If philosophers of mind aim to respect real physics rather than A-level chemistry (a concern previously directed toward metaphysicians \cite[p. 24]{LadymanRoss}), then  it is necessary to notice that in view of Noether's first theorem and its converse, physics implies a biconditional rather than categorical status of conservation laws.  Hence the traditional Leibniz-to-Dennett energy conservation objection fails.  One of course can and should employ neuroscience \emph{a posteriori} for more serious objections to interactionist dualism, but that is a completely different argument, one having only accidental connections to conservation laws.

The  failure of the Leibniz-to-Dennett energy conservation objection is not really news to philosophers of physics.  As Jeremy Butterfield wrote two decades ago:
\begin{quote} 
\ldots [A] traditional argument against interactionism is flawed.\ldots   The idea is that any causal interaction between mind and matter would violate the principle of the conservation of energy.\ldots  But, says the argument, physics tells us that energy is conserved in the sense that the energy of an isolated system is constant, neither increasing nor decreasing.\ldots  And there is no evidence of such energy gains or losses in brains. So much the worse, it seems, for interactionism. (Though traditional, the argument is still current; for example, Dennett endorses it (1991, pp. 34-35).) 

This argument is flawed, for two reasons. The first reason is obvious: who knows how small, or in some other way hard to measure, these energy gains or losses in brains might be? Agreed, this reason is weak: clearly, the onus is on the interactionist to argue that they could be small, and indeed are likely to be small. But the second reason is more interesting, and returns us to the danger of assuming that physics is cumulative. Namely: the principle of the conservation of energy is not
sacrosanct. \ldots [A]lthough no violations have been established hitherto, it has been seriously questioned on several occasions. It was questioned twice at the inception of quantum theory\ldots. And furthermore, it is not obeyed by a current proposal \ldots for solving quantum theory's measurement problem.  

In short: physicalists need to be wary of bad reasons to think physicalism is true, arising from naivety about physics.  \cite{ButterfieldPsychophysics}  \end{quote} 
Note that Butterfield, who is not himself sympathetic to dualism, doesn't seriously entertain the idea, common in the philosophy of mind, that failure of exact conservation constitutes an interesting objection.  Of course conservation fails, given that it follows from symmetry and the symmetry is violated by interactionism.  (The symmetry-to-conservation law link in Noether's theorem is common knowledge among philosophers of physics, though the converse is less well known.)  A more  interesting question for interactionists is whether their view violates conservation laws badly enough to run afoul of experiment.   
 Unfortunately Butterfield's insights have not had much effect on the philosophy of mind literature.

As discussed in the previous paper \cite{EnergyMental}, the  failure of the traditional Leibnizian objection  occurs in the most conservation-friendly territory available, classical local field theory with the principle of least action. Aspects of conservation laws in modern physics that have been largely overlooked in the philosophy of mind literature include some mentioned above and others as well:   locality (energy conservation holds not primarily for the whole universe, but in every place separately) \cite[ch. 5]{LangePhilPhys}, conditionality upon the absence of external influences, indeed biconditionality (symmetries and conservation laws being mutually entailing by Noether's first theorem and its converse \cite{Noether,BradingDissertation,BrownHollandNoether1,KosmannSchwarzbachNoether,ConservationSymmetryRomero}), and gentle failure (robustness implying that a small violation of conservation would not be catastrophic, in line with (\cite{ButterfieldPsychophysics}) and contrary to (\cite{Bunge}).  Basically the same holds for the conservation of momentum.  
Hence any quantum-based replies to the objection that one might give (as many do) push at an already open door (unless General Relativity closes the door), though possibly it doesn't hurt to have additional replies anyway, especially for a widely dismissed view.  Of course anyone aiming to give a \emph{positive account or a defense} of traditional mind-body interaction (which I do not aim to do) might need to take into account quantum physics somehow.  Given that this paper explains how General Relativity implies a less trivial objection than the traditional Leibnizian one, quantum replies could become relevant for that reason also.

The mind-body problem was of considerable interest to Herbert Feigl \cite{FeiglMindBody,FeiglMindBodyLogicalEmpiricism,HeidelbergerFechner}.  Feigl took the conservation of energy to be only empirically valid and potentially subject to refutation \cite[p. 472]{FeiglMindBody}, so he (wisely) would be unlikely to make the Leibnizian energy conservation objection to interactionism. Presumably he and other logical empiricists would  have been interested in the bearing of General Relativity on the  philosophy of mind (if there is one)  and likely would have been pleased to find therein some evidence against interactionism.   


\subsection{Enter General Relativity} 

What difference does General Relativity make? While General Relativity is by now over a century old, there seems to have been no correct exploration of its bearing on the conservation law issue in the philosophy of mind until now.   The previous paper closed with a mention that General Relativity makes a difference, and not for the better for the Cartesian \cite{EnergyMental}, contrary to claims in the miniscule literature that addresses the question at all, but left to this present paper the task of explaining and defending this claim.  Typically the energy conservation objection is based on a rather elementary grasp of physics, roughly high-school chemistry level, corresponding perhaps the physics of the 1860s, so any attempt to address the difference made by General Relativity is welcome. 

 Two better-informed authors have claimed  that conservation laws already fail in General Relativity even apart from dualism \cite{MohrhoffInteractionism,CollinsEnergy,CollinsSoulHypothesis}; thus there is no  conservation law left for dualism to ruin, so the usual objection is eliminated. Thus they invoke not a Noether biconditional relation between symmetries and conservation, but  General Relativity's supposed lack of conservation laws to respond to the Leibniz-to-Dennett objection.   The Mohrhoff-Collins invocation of General Relativity is, doubtless, a rhetorically impressive move:  by knowing much more physics than usual, one supposedly shows that the usual, supposedly scientific objection, fails.  Surprisingly to those outside the field of General Relativity, it is unusual to take the formal conservation laws and gravitational energy seriously in General Relativity.  The Mohrhoff-Collins proposal  takes that widely shared idea in the General Relativity literature and exploits it for the philosophy of mind.\footnote{Strictly speaking there is still a little conservation law to violate for some models, namely, constancy of total mass-energy in asymptotically flat space-times, with a localized matter distribution \cite[chs. 19, 20]{MTW}.  Such a distribution is contrary to the usual cosmological models, though not necessarily with the facts if one is willing to suppose that the cosmos differs enough past where we can see from how it is around here   \cite[p. 166]{Feynman} \cite{KleinIsland,SmollerTemple}. Of course this conclusion might need qualification in light of the apparently nonzero value of the cosmological constant $\Lambda.$ } %
Relatedly, Roger Penrose has proposed that the energy non-conservation difficulty of GRW spontaneous collapse quantum theory might find  resolution in some future framework through linkage to the gravitational energy nonlocalization \cite[pp. 334, 344-346]{PenroseShadows}; this proposal also has links to the philosophy of mind.

Generally there is  something of an anti-realist tendency regarding energy conservation in General Relativity, at least regarding the local conservation laws, the kind that elsewhere  modern physics usually employs.  An old standard textbook decreed that ``[a]nybody who looks for a magic formula for `local gravitational energy-momentum' is looking for the right answer to the wrong question.'' \cite[p. 467]{MTW}
If gravitational energy isn't really anywhere in particular, then non-gravitational energy doesn't satisfy an honest conservation law, either, because of interconversion between material energy and gravitational (pseudo-?)energy.  Instead one has only a balance law $\nabla_{\mu} T^{\mu\nu}=0$ describing how non-gravitational energy fails to be conserved due to gravitational influence.  This balance equation typically does not imply any global conservation law \cite[pp. 236, 269-271]{WeylSTM} \cite[p. 280]{Landau} \cite[p. 465]{MTW} \cite[p. 139]{LordTensors} \cite[p. 141]{Stephani}.
While most physicists now accept that gravitational energy exists (which has not always been the case \cite{KennefickWaves}),  it is difficult to make sense of this nonlocalizability, an apparent lack of any objective location.  In recent decades some philosophers have sought conceptual clarity by denying the existence of  gravitational energy  \cite{Hoefer,DuerrFantasticBeasts}.  

But if one is a realist about gravitational energy localization and conservation, one will be quite  disinclined to accept the Mohrhoff-Collins move. 
 General Relativity has uncountably infinitely many symmetries of the action\footnote{It is quite common to introduce a red herring by bringing up symmetries (or the lack thereof) of the \emph{geometry}, as though a spatio-temporally varying metric were relevant to the existence of conservation laws \cite{CarrollEnergyIsNotConserved,MotlEnergyNotConserved,HossenfelderEnergy,SiegelEnergy}.  But Noether's theorem does not know or care about geometry; it carries only about symmetries of the action \cite{Noether} or, if one permits nonvariational fields, whether the nonvariational fields have symmetries \cite{TrautmanUspekhi}.  To ask for a field with Euler-Lagrange equations to have symmetries as well is to require  supererogation.} and thus, at least formally, infinitely many conserved currents \cite{BergmannConservation}.  The gravitational energy realist takes this mathematics seriously and infers that General Relativity is \emph{more} conserving of energy than other theories, not less so, so broadly Cartesian mental causation should be harder, not easier, given General Relativity \cite{EnergyGravity}. The fact that Einstein's equations \emph{alone}, without help from the matter field equations, entail conservation laws is further evidence that General Relativity is unusually supportive of conservation laws rather than unusually lax as is often claimed.  But is one left with a choice of a common  non-realist interpretation and a  less popular realist interpretation, with no means of definitive adjudication for the consequences for mental causation other than sociology?

Fortunately there \emph{is} a way to resolve the question of the bearing on mental causation objectively, one of the sort that Leibniz himself would have approved: ``Calculemus!''  One can do new calculations involving the generalized Bianchi identities, without taking  a prior interpretive stand on the controversies involving gravitational energy.  These new calculations, being tensorial, are entirely free of the confusion and controversy that have swirled about gravitational energy for a century.  When one does the generalized Bianchi identity mathematics, one finds that General Relativity makes Cartesian mental causation \emph{harder}, not easier.   Thus one can go beyond interpretive stances and achieve objective results.

In the simplest case, when the Cartesian mental influence is a scalar field (a single number at each point, the same in all coordinate systems), that influence must be spatio-temporally constant, as will appear below.  Spatio-temporal constancy is obviously absurd for the influence of a human  mind unless that influence is zero, so that there is no Cartesian mental causation after all. Unlike the traditional Leibnizian objection, this new objection from General Relativity does not assume something equivalent (given Noether's theorem and its converse) to non-interaction in order to infer non-interaction.  Thus General Relativity resists external influences by trying to force them to vanish.  While the simplest toy model of mind-body influence yields a negative result for the ability of the non-physical mind to influence the body, this result weakens as the complexity of the model arises. The complexity of the model for mental causation is also bound up with larger world-view issues, as will appear below.  
Because the majority (non-localization) interpretation of gravitational energy  has the wrong heuristic force, that non-localization interpretation is somewhat undermined by the objective Bianchi identity results.  Correspondingly, the minority  realist interpretation has the correct heuristic force and so is  somewhat confirmed. 

Realism about gravitational energy is philosophically interesting for another reason, namely, the relevance to conserved quantity theories of causation \cite{FairCauseFlowEnergy,DowePhysicalCausation}.  A number of authors have recognized that the conventional doubts about the reality of localized gravitational energy and the consequent lack of true conservation laws pose problems for regarding energy and momentum as conserved quantities, thus depriving conserved quantity theories of causation of their star examples \cite{RuegerCauseEnergy,DoweDefended,CurielCauseGR,LamCauseConservedGR}.  Realism about gravitational energy localization and hence the pseudotensorial local conservation laws would imply that General Relativity is no longer an objection to theories of causation that take energy to be a conserved quantity.  One should, however, get accustomed to making energy plural (conserved quantit\emph{ies} theory of causation), because the  mathematics speaks of infinitely many conserved energies and momenta if it speaks of any.  One might see the help given to conserved quantity theories of causation as another theoretical virtue that counts somewhat in favor of realism about gravitational energy(s). 


\section{Mental Causation:  Carroll's Foundling Program} 

The question whether the causal influence of immaterial minds (or for that matter, nonredundantly causal mental properties) can be modelled in physics, though not novel, has become the more timely in light of the recent interventions of Sean Carroll, Caltech cosmologist, amateur philosopher, and proponent of ``poetic naturalism,'' which one can choose as one's religion at Facebook \cite{CarrollFacebookReligion}.  Judging by Carroll's rhetoric, the answer is clearly negative, apparently apart from calculations, literature searches, or experiments.   
Nonphysical mental causation would ``overthrow everything we think we have learned about modern physics'' \cite{CarrollMindEnergy}.  Readers  versed in physics might be startled to hear that, say, the spin-statistics theorem (that one should quantize bosonic fields using commutators and fermionic fields using anticommutators, a standard result of quantum field theory (\emph{e.g.}, \cite[p. 88]{Kaku} \cite[pp. 52-56]{PeskinSchroeder}), or the apparent empirical adequacy of representing all spatio-temporal-gravitational physics in terms of a single space-time metric tensor, would be overthrown by immaterial souls with causal influence.  Lay readers, on the other hand, might be inclined to accept Carroll's claim.  

Carroll's recent semi-popular philosophy book  \cite[pp. 212, 435-441]{CarrollBigPicture}, which purports to derive science-based conclusions for great questions including the philosophy of mind and the question of life after death, %
has been  reviewed in \emph{Science}.  Where the book discusses the bearing of physics on the philosophy of mind, it continues the earlier program.  Carroll's %
 rhetorical questions can serve to inspire  a genuine inquiry into his questions.  It turns that whereas the traditional Leibnizian energy conservation objection is question-begging \cite{EnergyMental}, Carroll raises a good question in asking how, if at all, physics could be modified so as to take account of the influence of immaterial minds on the electrons in brains.  In teaching a wide audience about  ``Physics and the Immortality of the Soul,''  Carroll displays  the Dirac equation, though he later claims that the mere existence of the equation, not its detailed form, matters.  His informal argument against dualism and immortality follows.  
\begin{quote}    
As far as every experiment ever done is concerned, this equation is the \emph{correct} description of
how electrons behave at everyday energies. It's not a complete description; we haven't included
the weak nuclear force, or couplings to hypothetical particles like the Higgs boson. But that's
okay, since those are only important at high energies and/or short distances, very far from the
regime of relevance to the human brain.

If you believe in an immaterial soul that interacts with our bodies, you need to believe that this
equation \emph{is not right}, even at everyday energies. There needs to be a new term (at minimum)
on the right, representing how the soul interacts with electrons. (If that term doesn't exist,
electrons will just go on their way as if there weren't any soul at all, and then what's the point?)
So any respectable scientist who took this idea seriously would be asking - what form does
that interaction take? Is it local in spacetime? Does the soul respect gauge invariance and
Lorentz invariance? Does the soul have a Hamiltonian? Do the interactions preserve unitarity
and conservation of information?

Nobody ever asks these questions out loud, possibly because of how silly they sound. Once you
start asking them, the choice you are faced with becomes clear: either overthrow everything we
think we have learned about modern physics, or distrust the stew of religious
accounts/unreliable testimony/wishful thinking that makes people believe in the possibility of
life after death. It's not a difficult decision, as scientific theory-choice goes.
\cite[emphasis in the original]{CarrollMindEnergy}
\end{quote} 

The careful reader will notice that Carroll looks for evidence only to experiments aimed at testing how electrons, electromagnetism and gravity interact, ignoring the possibility that a new term on the right could vanish in the apparatus at SLAC, Fermilab, CERN and the JINR, while being nonzero elsewhere, such as in brains  \cite{ThompsonMind}, or perhaps on occasion in monasteries, churches, ashrams, seances, or the like --- places that spirits, if they exist and act on matter, might find especially relevant. To my knowledge most interactionist dualists think that souls act on brains, or perhaps on bodies as a whole, but not on the rest of the physical world.  Proponents of  psychokinesis, rare but not nonexistent among parapsychologists, might demur from this limitation \cite{BraudeLimitsInfluence,LevitationGrossoCopertino}; but such claims are purportedly scientific and not paradigmatically  religious or traditional dualist claims. 
Carroll likewise evidently envisages that psychic powers, souls, \emph{etc.} must be envisaged as a sort of subtle physics, implemented \emph{via} new physical particles/fields or the like.  There are, of course, people who believe precisely this. 
One can find views in this neighborhood in some of the most visible work in parapsychology \cite{ParapsychologyHandbook21st}.   I largely agree with Carroll's critique: if there were physical fields/particles that mediated psychic powers, experimental physics ought to have discovered them, but it hasn't, so are almost certainly not such things.

The problem is that Carroll claims to have shown much more, not merely that there isn't some subtle physics explaining paranormal powers, but that there are no souls acting on bodies and (therefore?)  no life after death  \cite[p. 4, chs. 19-24]{CarrollBigPicture}.  Yet his critique entirely fails if such events, powers, \emph{etc.} are implemented \emph{via} non-physical personal entities:  immaterial souls, God, angels, demons, genies, ghosts, or the like.  Being non-physical and personal, such influences are not equally available and testable everywhere and always; rather, they are available and testable (if they exist) only in particular space-time regions---in living brains, around saints, in prayer meetings, at seances, 
near witchdoctors \cite{AnthropologyBeingChanged}, or the like.  Brains, at least, are rather reliably available, but still sufficiently localized that experiments on thigh muscles or in nuclear physics laboratories are irrelevant (a point overlooked surprisingly often).  By extrapolating from experimental physics and ignoring the potential disanalogies introduced by the spiritual context (at least as judged by some of the people  whose views he claims to refute), Carroll  has generated a conflict between such testimonies and experimental physics.  Given that he is on the offensive dialectically, his argument  begs the question.   

The point that evidence for anomalous events needs to be sought \emph{where and when such events allegedly happen(ed)}, not where one finds it convenient to look, has been made before, such as in the 18th century by defenders of miracles Gottfried Less, George Campbell and William Paley
\cite{CraigMiraclesGerman}. Otherwise one is like the proverbial  intoxicated man who looks for his keys under the street light rather than where he probably dropped them.  Miracles and soul-to-body causation, if they happen at all, are not uniformly  distributed events.  Evidence that such events have occurred in certain times and places does not contradict evidence that they haven't occurred elsewhere.  If there is evidence that such events have happened, then there is evidence that the allegedly universal law isn't a universal law, and there is no conflict between the evidence for the particular exception and the supposedly stronger evidence for the universal law.  Perhaps Hume's critique of miracles underlies Carroll's reasoning, but that is hardly an uncontested argument \cite{EarmanHume}.  

Readers versed in  philosophy  might think that an argument from the (supposed) failure of interactionist dualism to the denial of an afterlife is a bit quick; one might consider such  presently less fashionable substance dualist options as occasionalism, pre-established harmony, and epiphenomenalism before discarding an afterlife on the basis of troubles with interactionism.  Given that my purpose here is not to defend ideas of an afterlife but to develop Carroll's abandoned program of evaluating interactionism, I will not pursue such questions further, however.

Carroll's claims about the  bearing of contemporary physics on interactionism are controversial.  
\begin{quote} 
How an immaterial soul might interact with the physical body remains a challenging question for dualists even today, and indeed it has grown enormously more difficult to see how it might be addressed [compared to the days when Elisabeth queried Descartes].\ldots   \cite[pp. 212]{CarrollBigPicture}, 
\end{quote}  
But in fact a fair number of serious people  have suggested that such interaction is in fact made \emph{easier} rather than harder, due to quantum mechanics \cite{EcclesSelfBrain,Atmanspacher}. Wigner took quantum mechanics to provide reason to infer mental influence on the physical world \cite{WignerMindBody}.  Such claims are not merely from decades ago; there are recent proposals from Princeton's Hans Halvorson \cite{HalvorsonSoulHypothesis} and Cambridge's Adrian Kent \cite{KentQuantumQualia}.  I set such claims aside, however, because I am interested in ways that modern physics might make such mental conservation harder rather than easier.

One might be tempted to take Carroll's claim to be akin to those of (mostly) physicalist philosophers of mind who raise the energy conservation issue, but in fact there are key differences. For one, Carroll doesn't bring up energy conservation (at least not here, though eventually he does elsewhere \cite[p. 356]{CarrollBigPicture}). Perhaps for a theoretical physicist, energy conservation would be  too obviously question-begging an objection to mention, given the connection between symmetries and conservation laws?  Or perhaps Carroll's interpreting General Relativity as not having  energy  and momentum conservation laws \cite{CarrollEnergyIsNotConserved,CarrollSpacetimeGeometry} explains his failure to list it as a casualty of soulish causal influence.  
Elsewhere he claims that the total energy of the universe is $0$ in light of General Relativity \cite[p. 201]{CarrollBigPicture}, a claim that seems to backpeddle from that strong anti-realism.

 Carroll   performs dualists a service by clearly formulating the issue and sketching a potentially illuminating  research program.   Indeed much of the \emph{philosophical} worry about interactionism, such as Princess Elisabeth's understandable inability to conceive how such disparate entities as mind \emph{with no spatial location} and matter could interact, already starts to dissipate upon being framed in contemporary physics.  While it has sometimes been unclear what this objection is at least in modern philosophical reiterations \cite{KimLonelySouls,BurgeMind,CartesianPsychophysics}, formulation in modern physics makes the objection even less intelligible.  Interactions happen by terms in equations of motion;  two fields  interact, such as the electromagnetic field and the electron field, if their  Lagrangian density a term involving products of both fields (and/or their derivatives), so that each field appears in the other's equation of motion \cite{Kaku,PeskinSchroeder}. This fact, as it happens, makes something that naively seemed highly implausible --- that something could be real and yet invisible (due to not coupling to the electromagnetic field) --- now quite pedestrian, as the search for dark matter highlights.  Likewise there is no obvious difficulty in having two objects occupy the same space, if one is made of one collection of fields, the other is made of another collection of fields, and the two collections of fields have no mutual interaction terms in the Lagrangian. (So it is at least in \emph{classical} field theory.)  What, then, is to stop an immaterial substance from having effects that appear in a Lagrangian?  At seems that there is just nothing much else to say in response to ``how?''  unless  one rejects the usual standards of modern  physics. The de-materialization of everything including matter into fields, suggested around 1930 by Pascual Jordan and by now generally accepted in physics, has left the contemporary understanding of the physical world far more ethereal (so to speak), ghostly, tenuous, and mathematical than either the 17th century mechanical philosophy or tables-and-chairs daily experience would have suggested.  Carroll is not worried about spatially non-located souls, because he wants to refute the following view:
\begin{quote} 
Very roughly speaking, when most people think about an immaterial soul that persists after
death, they have in mind some sort of blob of spirit energy that takes up residence near our
brain, and drives around our body like a soccer mom driving an SUV. \cite{CarrollMindEnergy}
\end{quote} 
Hence the greatest source of difficulty faced by Princess Elisabeth, nonspatiality of souls, does not concern Carroll.  After conceiving an interesting research program, the implementation of which might in fact tend to undermine interactionist dualism, Carroll abandoned it.  I, however, will take it up, at least in part, thereby finding a new objection to interactionist dualism from General Relativity.  Carroll's  foundling research program suggests exploring what could possibly be ``new term (at minimum) on the right, representing how the soul interacts with electrons'' to assess the prospects for modeling soul-brain interaction in a fashion congenial to theoretical physicists.  Such a term would, if all goes well for the dualist, preserve core physical principles insofar as one could reasonably expect (such as locality, gauge invariance, \emph{formal} Lorentz invariance (though obviously violating all the symmetries \emph{de facto} in those spatio-temporal regions where the soul acts on the brain), and  unitarity (conserved positive normalized probabilities).  I aim for something less ambitious, namely to see what bearing General Relativity has on the energy conservation question.  Happily for Carroll and unhappily for interactionist dualists, it turns out that there \emph{is} a difficulty in the vicinity.

To try to write down a local field interaction broadly consonant with modern physics, it is natural to treat souls as spatially located and extended,  denying some aspects of the Cartesian view  asserted by Foster \cite{FosterSoulBody} and criticized by Fodor \cite{FodorMindBody}.
Spatially located souls are both mildly fashionable nowadays (at least as souls go \cite{EcclesMindwaves,HaskerCorcoran,ZimmermanDualism}) and closer to  the historical mainstream than one might have thought  \cite{LockeSolidSouls,ZimmermanDualism,ReidSpatialSpirits,ClarkeExtendedSoul}.  ``For even if philosophers today very often take for granted that immaterial entities have no location, this is in fact quite an extraordinary view, historically speaking.'' \cite[p. 328]{Pasnau}
  Thus one naturally  doubts one of Fodor's assumptions about dualism: ``If the mind is nonphysical, it has no position in physical space. How, then, can a mental cause give rise to a behavioral effect that has a position in space?'' \cite{FodorMindBody}
 Indeed Lycan has advised dualists to give up nonspatiality \cite{LycanDualism}:
\begin{quote} \vspace{-.07in} 
The big problem for interaction is and remains the utter nonspatiality of Cartesian egos. (By now we can all tolerate action at a distance. But action at a distance is at least at a distance.) My suggestion is that the dualist give up nonspatiality. Descartes had his own 17th-century metaphysical reasons for insisting that minds are entirely nonspatial, but we need not accept those. Why not suppose that minds are located where it feels as if they are located, in the head behind the eyes? [footnote  suppressed] \cite{LycanDualism} \vspace{-.07in} 
\end{quote}
Spatial extension also leaves more room for efforts toward detailed dualist neuroscientific proposals, such as  Eccles made \cite{PopperEccles,EcclesSelfBrain,CartesianEccles}, some of which have been criticized \cite{ClarkeSoul}. 
 Jaegwon Kim has briefly entertained the idea that spatially locating souls might help to address his pairing problem \cite[section V]{KimLonelySouls}.  %


\section{Quantum Field Theory in Curved Space-time as  a Precedent}

The subject of quantum field theory in curved space-time provides a relevant precedent for assessing Carroll's  claim that soul-to-body influence would  ``overthrow everything we think we have learned about modern physics\ldots.''  He elsewhere says:
\begin{quote} 
The Core Theory of contemporary physics describes the atoms and forces that constitute our brains and bodies in exquisite detail, in terms of a rigid and unforgiving set of formal equations that leaves no wiggle room for intervention by nonmaterial influences.'' \cite[p. 212]{CarrollBigPicture} 
\end{quote} 
It is difficult to know what such wiggle room would be such that modern physics lacks it.  Is the problem that modern physics is deterministic?  Of course it isn't clear that modern physics is deterministic---that depends on one's interpretation of quantum mechanics---but it is no more deterministic than Newton physics.  But even deterministic physics as such presents no obstacle; if there are causally active souls, presumably they introduce novel forces into the laws and make the equations different from what they would have been without causally active souls.  That the equations of motion in the absence of soulish influence are deterministic is irrelevant, because those equations would not apply to the physical world in case souls act.  

What about rigid and unforgiving equations?  Physicists write  books on how to do quantum field theory in the presence of spatio-temporally varying influence:  quantum field theory in curved space-time   \cite{BirrellDavies,Fulling,WaldQFTiCST,DeWittGlobalQFT,vanNPathCurved,ParkerQFTiCST,HackQFT},  and the answer is not that everything goes splat and all physical knowledge is lost.\footnote{Duff's critique  \cite{DuffQFTinCSTinconsistent} in terms of the  failure of the expected equivalences under field redefinitions on account of mixing quantum and classical physics (each of which separately supports such equivalences) is interesting.  The question of field redefinitions is considered in more detail below.  }   
  The soul's influence, in varying spatio-temporally and lacking symmetries, is not altogether unlike the influence of curved space-time,  a sort of external potential, on quantum fields.  Indeed one might think that souls' influence would be far less disruptive in some respects, because souls' influence would only violate time- and space-translation symmetries in the few modest spatio-temporal regions where souls allegedly act, while leaving intact enough idea of the action of the Poincar\'{e} group to define particles, for example.  One might hazard some conjectural answers to Carroll's rhetorical questions:  ``what form does that interaction take? Is it local in spacetime?''  Presumably so to the latter question.  ``Does the soul respect gauge invariance and Lorentz invariance?''  Presumably, yes and (excepting the obvious sense of violating the translation symmetries in some localized spatio-temporal regions) yes.  ``Does the soul have a Hamiltonian?'' No, but its \emph{effects} should   figure in a modified action principle for physical fields.  ``Do the interactions preserve unitarity and conservation of information?''  Presumably they do as much as physics needs. 
(He provides an additional list of useful questions in the book \cite[p. 216]{CarrollBigPicture}.)  Carroll's claim that souls acting on bodies would ``overthrow everything we think we have learned about modern physics'' \cite{CarrollMindEnergy} is epistemically possible.  However, such a conclusion should be not a bare assertion  for mass consumption, but the result of detailed consideration of similarities and dissimilarities between soulish influence and gravitational influence in the context of quantum field theory in curved space-time, an extant theory of quantum fields interacting with an asymmetrical external potential.  The next section shows how one  \emph{can} pursue analogously detailed questions in the context of classical gravity and how the results are somewhat congenial to Carroll's poetic naturalist religious project, at least if one supplements or replaces his treatment of conservation laws in General Relativity.  In General Relativity one does find a somewhat rigid and unforgiving set of formal equations at last---although this rigidity is finite.  If one were attempting to grope towards biological realism, one would, as Carroll notes, need to have the soul interact especially with electromagnetism and electrons, but that is not my aim.


\section{Does General Relativity {Help} Mental Causation?}

According to Eddington, the Archbishop of Canterbury once asked Einstein about the relevance of relativistic physics for religion;  Einstein replied that relativity was a purely scientific theory and  had no relevance to religion \cite[p. 7]{EddingtonPhilPhysSci}.  Those who have suggested that General Relativity provides evidence for creation \emph{ex nihilo} might disagree with Einstein; doubts about such claims have been expressed \cite{PittsKalam}.  
But gravitational energy is another area where General Relativity might have some bearing on religion, whether for good or for ill.  As appeared above, a couple of authors have claimed that General Relativity makes it easier for souls to act on bodies \cite{MohrhoffInteractionism,CollinsEnergy,CollinsSoulHypothesis}.
Such a claim, if true, would make General Relativity positively relevant to religion.

However, an early result of attempting to carry out Carroll's foundling program to model the causal  influence of immaterial minds on the material world turns out to yield a novel \emph{objection} to such causal influence from General Relativity.  This objection naturally involves delving into the topic of gravitational energy and general relativistic conservation laws.  General  relativistic conservation laws have been a thicket of controversy for a century, though philosophers have rarely discussed the subject until recently \cite[p. 86]{RussellMatter} \cite{GravesEnergy,RuegerCauseEnergy,CurielCauseGR,Hoefer,CollinsEnergy,EnergyGravity,LamCauseConservedGR,LamEnergy,CollinsSoulHypothesis,DewarWeatherallGravitationalEnergyNewton,ReadEnergy,DuerrFantasticBeasts}.

 Mohrhoff and  Collins (independently) follow a very common  interpretation in the physics and philosophy of physics communities \cite[p. 467]{MTW} that denies the physical meaning of the formal mathematics of ``pseudotensor'' conservation laws in General Relativity.  The orthodox view in  General Relativity for some time has been, to recall from above, that ``[a]nybody who looks for a magic formula for `local gravitational energy-momentum' is looking for the right answer to the wrong question.'' \cite[p. 467]{MTW}  Mohrhoff and Collins, unlike others writing in the philosophy of mind, were aware of this standard view about General Relativity.  One can compare their comparatively well informed view to what was intended as an uncontroversial remark deferential to standard physics of  conservation laws: 
\begin{quote}
\ldots  about as well established as anything could be in physics, the conservation of mass and energy tells us that in a ``closed'' system changing over time, the net total of mass or energy in the system, stays the same. \cite[p. 41]{Westphal}
\end{quote}
By contrast Mohrhoff and Collins knew the situation in the General Relativity literature, and then proceeded to take it seriously in the philosophy of mind, thus finding  a loophole useful for dualists  \cite{MohrhoffInteractionism,CollinsEnergy,CollinsSoulHypothesis}. The issue with gravitational energy is that gravitational energy-momentum depends in a peculiar and \emph{essential} way on the choice of coordinate description, which presumably physically real quantities would not do. That is, gravitational energy-momentum is not described by a ``geometric object'' \cite{BergmannConservation}, a set of components relative to each local coordinate system and a transformation rule relating coordinate systems where both apply \cite{Schouten,Nijenhuis}.  One can have energy absent when intuitively it ought to be present (as Schr\"{o}dinger noted) or energy present when intuitively it ought to be absent (as Bauer noted) \cite{SchrodingerEnergy,BauerEnergy,Cattani}, at least using Einstein's canonical pseudotensor, and presumably using the other pseudotensors as well.  The prolonged absence of any sensible physical interpretation of the general relativistic conservation laws has encouraged a widespread view that any local notions of conservation or spatial presence of gravitational energies lack objective physical significance. 
There was even a time in the 1930s and 1950s when some leading general relativists (at times including Einstein) doubted the existence of gravitational waves and/or gravitational energy altogether, although this view largely disappeared due to the late 1950s sticky bead argument that gravitational wave energy could be converted to heat \cite{BondiNature,FeynmanUnpublished,KennefickWaves}.  (Recently the denial of gravitational energy has been proposed  once more \cite{Hoefer,DuerrFantasticBeasts}. Relatedly, Cooperstock has long argued that there is no gravitational energy outside matter and hence no energy in gravitational waves  \cite{CooperstockEnergy}.)   Hence there are no local conservation laws, notwithstanding (it is said) the impression given by local pseudotensor conservation laws.  (There is some resistance to such claims, however, including new ideas for dealing with the two main problems, the non-uniqueness  \cite{NesterQuasiPseudo} and coordinate dependence \cite{EnergyGravity} of pseudotensors; in both cases there is  suggestion that multiplicity is a meaningful virtue rather than a defect.  Peter Bergmann once indicated that, roughly speaking, these two infinities are in fact one and the same infinity \cite{BergmannConservation}.)




\section{Local Conservation in General Relativity?}

 One reason  that General Relativity makes the mind-body problem harder  builds on the mathematical fact that  General Relativity not only entails local conservation laws (as do other theories, though they require more help than General Relativity does), but also (unusually) is entailed by conservation laws \cite{Noether,BergmannConservation,Anderson,EnergyGravity,ConverseHilbertian}.  Descartes's effort to found physics on conservation laws \cite{BelotConservation} is finally realized. The mathematics is old and uncontroversial, even if somewhat forgotten. 

 My view is distinctive in proposing that this mathematics can be taken seriously, that is, realistically.  (Nester and collaborators make somewhat similar claims   \cite{NesterQuasiPseudo}, though they are mostly interested in interpreting the nonuniqueness of pseudotensors and in pointing out that pseudotensors, considered bad and pass\'{e}, are quasi-local, a property considered good and modern. Nester elsewhere admits to not following the advice of his teacher, one of the authors of (\cite[p. 467]{MTW}) \cite{NesterQuasi}.) General Relativity has uncountably infinitely many time translation symmetries of its Lagrangian \cite{BergmannConservation}:  one definition of being at rest (part of a vector field with components $(1,0,0,0)$ in some coordinates, at least in a neighborhood) is equivalent to another definition of moving in an irregular way (that same vector field with  wiggly components in some other coordinate system).   By Noether's first theorem, such a symmetry as time translation (which the first definition exemplifies) implies a conservation law (of energy). But this vector field is itself arbitrary, apart from being non-zero and time-like (at least in a neighborhood).  Thus formally one has infinitely many conserved energies. This claim is rarely considered (but see \cite{SzabadosSparlingHAS}) and is considered paradoxical if it is considered at all.  I propose that the paradox is  due to a tacit and unargued Highlander-type assumption that ``there can  be only one'' energy.  Why can't all the energies be real \cite{EnergyGravity,EnergyGravityConf}? It is a bit odd to suggest that a theory logically equivalent to infinitely many (formal) conservation laws, which exist thanks to infinitely many translation symmetries of the Lagrangian (in light of Noether's theorem), in fact lacks any conservation laws at all, in effect setting  $\infty=0$.  Just because it's infinite  doesn't mean it's $0$ (to recall a phrase used around quantum field theory in  the late 1940s).  If one embraces infinitely many energies, then the Noether symmetry-conservation link resurfaces and the counting works:  $\infty=\infty$. 

A linguistic analogy will illustrate why pseudotensorial behavior could be a virtue.  Suppose that one is learning about Mary/Mar\'{i}a and encounters the English sentence ``Mary is short'' and the Spanish sentence ``Mar\'{i}a es alta''  and tries to relate them by translation.  This task is frustrated, because ``Mar\'{i}a es alta'' translates to ``Mary is tall.'' How can one and the same Mary/Mar\'{i}a  be both short and tall?  Perhaps Mary/Mar\'{i}a is a person who lacks any objective height?  This proposal is analogous to the majority view about gravitational energy:  one builds the pseudotensor using the same functional form with the metric tensor components related using the tensor transformation law and finds the pseudotensor ascribing rival gravitational energy properties to the same points relative to different coordinate systems.   
 When Mary and Mar\'{i}a walk into the room together, the difficulty vanishes:  $Mary \neq Mar\acute{\i}a$, so why should two different people have the same properties?  Instead of a description of one person in two languages, one had a description of two different people, each in her own native language. Languages, like coordinate systems, are arbitrary conventional ways to express realities that transcend those specific conventions; translation is like the transformation rule for a tensor or other geometric object. 
Much as people have native languages (typically unique), conserved energies  have native coordinate systems, such as those in which the corresponding translation  vector takes the form $\xi^{\mu} =(1,0,0,0),$ which allows one to pull the vector components out of the natural conserved current equation $\partial_{\mu} (\xi^{\nu} \mathcal{T}^{\mu}_{\nu})=0$ (or some generalization thereof \cite{SorkinStress}) to get a more coordinate-dependent equation for a conserved rank {\bf $2$} expression  $ \xi^{\nu} \partial_{\mu} \mathcal{T}^{\mu}_{\nu}=0$ (all indices being in their  primordial locations).  The conventional tacit assumption in General Relativity has been that there is only one energy (or four energy-momenta) for a gravitational energy-momentum pseudotensor to describe; thus the one energy (or four energy-momenta, rather) must be described in all coordinate systems, albeit with frustratingly incompatible properties.  But the assumption that there is (at most) only one energy 
 is motivated largely by lack of imagination \cite{EnergyGravity,EnergyGravityConf}
or perhaps the wish for simple bookkeeping.  If one uses Noether-inspired bookkeeping, one would expect infinitely symmetries to yield infinitely many momenta.  If a pseudotensor describes infinitely many energies, each with its own preferred coordinate system, then pseudotensorial behavior becomes intelligible.

 Indeed one can go further:  pseudotensorial behavior becomes inevitable and virtuous, because a geometric object with finitely many components (presumably $10$ or $16$) could only describe one energy, its components in each coordinate system all being merely faces of one and the same conserved quantity.  Pseudotensoriality permits something that would be impossible otherwise:  it enables a finite-component object to describe infinitely many conserved quantities, which one expects to exist due to the infinity of symmetries. Hence pseudotensorial behavior is a virtue, not a vice, from this perspective.

If one takes these conservation laws seriously, then the situation for General Relativity, one might think, is basically the same as for other field theories; the dualist  still can (and must) appeal to the (bi)conditional nature of conservation laws to deflect the objection.  (Actually one needs to work harder than that!)  People who find it absurd that General Relativity should be a resource for dualists, as I do --- in my experience physicists tend to recoil from Collins's inference when confronted with it 
 --- could take that judgment as a motivation to take  gravitational energy-momentum pseudotensor conservation laws more seriously.  Doing so would also leave more  room for conserved quantity theories of causation, as will be discussed again at the end.

While such ideas strike me as illuminating, the reader does not need to sympathize with them in order to profit from the covariant calculations below.  Those calculations, however, might reflect some plausibility back onto the realism about pseudotensorial energy that had the correct heuristic force.


\section{Bianchi Identities Constrain Mental Influence}

A more clearly compelling  reason  that General Relativity does not help the dualist is that one can show mathematically that whereas pre-GR theories will tolerate the introduction of an external mental potential(s) without objection (with symmetries and conservation laws both spoiled thereby in accord with Noether's first theorem and converse \cite{Noether}), General Relativity \emph{actively resists} such an external influence by striving to make it vanish.  One can, but need not, construe this point as a general relativistic energy conservation objection.  Carroll's undefended claim that nonphysical mental causation would ``overthrow everything we think we have learned about modern physics''  \cite{CarrollMindEnergy} cannot be defended (at least by him) in terms of energy conservation because he denies it \cite{CarrollEnergyIsNotConserved}; his General Relativity textbook does not apply Noether's theorem to General Relativity \cite{CarrollSpacetimeGeometry}.  However, his clearly framing the question of how to modify the  Dirac-Maxwell-Einstein equations to permit immaterial mental influence suggests a research question.  If one attempts to answer this question,  then a new difficulty for interactionist dualism  from the generalized Bianchi identities emerges.   If one introduces souls that act on physical fields in theories different from General Relativity, then some of the conservation equations will be false,\footnote{Whereas physicists happily enough utter expressions such as ``conservation laws fail'' or are ``violated,'' meaning that the continuity equation is not true,
philosophers tend to find such expressions paradoxical because most think that laws must be true and violations are naughty.  Thus I have attempted to avoid such expressions.} 
 because the souls act as sources/sinks for energy and momentum.  But in General Relativity,  the generalized Bianchi identities, related to Noether's \emph{second} theorem, are also relevant.  In this application one can show  that they tend to constrain how the soul could act on matter.   General Relativity presents dualism not a new solution as Mohrhoff and Collins envisaged, but a new problem  of surprising technical intricacy.   %

	It is widely accepted in computer science and various parts of engineering that  one can understand something better by trying to break it:  destructive testing.  Popper's falsificationism is perhaps somewhat similar in spirit.  In any case General Relativity proves very different from earlier physical theories in that whereas earlier theories passively tolerate the introduction of external causal influences (applied fields, external potentials), General Relativity actively resists them by trying to force them to vanish.  One might take the view that laws of nature describe the tendencies of physical entities, while leaving unspecified whether any nonphysical entities influence the physical \cite{WachterCausalClosureImmaterialThings}. 
The Bianchi identities exhibit how and to what extent resistance  occurs.  

 Let the action for General Relativity with matter $u$ (indices suppressed) and influence $\Psi$ from immaterial minds  be $S[g_{\mu\nu},u, \Psi]$. While little can be said with any confidence about how the mental influence $\Psi$ enters in, let us treat it as a  field (or collection thereof) with some definite coordinate transformation properties, entering the action algebraically or with up to some finite number of derivatives, with $S[g_{\mu\nu},u, \Psi]$ invariant (at least up to a boundary term) under coordinate transformations as usual.  Crucially, $\Psi$ does \emph{not} have any equations of motion from the principle of least action:  one does not postulate the vanishing of $  \frac{ \delta S}{\delta \Psi}$.  
(One might assume the boundary condition that $\Psi=0$ outside brains, or wherever immaterial minds might act.)  Using coordinate transformations that are trivial at the boundary (to annihilate boundary terms), one has 
\begin{eqnarray}  \int d^4x \left( \frac{ \delta S}{\delta g_{\mu\nu}} \pounds_{\xi} g_{\mu\nu} + \frac{ \delta S}{\delta u} \pounds_{\xi} u +
\frac{ \delta S}{\delta \Psi} \pounds_{\xi} \Psi \right)=0.
\end{eqnarray}
Because of the matter field equations  $  \frac{ \delta S}{\delta u} = 0$, the second term vanishes. 
Because of the gravitational field equations $ \frac{ \delta S}{\delta g_{\mu\nu}}=0$ (Einstein's equations or some modification thereof), the first term vanishes.

To go further, it is necessary to be a bit more specific about the mental influence $\Psi.$
To emphasize the possibility of multiple components, one can write $\Psi$ as $\Psi^A,$ where the index $A$ runs over all the components.    Suppose (with little loss of generality) that under infinitesimal coordinate transformations along the vector field $\xi^{\mu},$  $\Psi^A$ transforms as $$\pounds_{\xi} \Psi^A = \xi^{\mu} \partial_{\mu} \Psi^A - C^{A\nu}_{\mu}(\Psi, g?) \partial_{\nu} \xi^{\mu} $$ \cite{AndersonPrimary}.  
This expression lacks a term independent of $\Psi^A$ and so suits the assumption above that $\Psi^A=0$ is physically distinguished as the state when the soul does nothing to the body.  
While  $C^{A\nu}_{\mu}(\Psi, g?) $ is presumably linear in $\Psi^A$, allowing the transformation rule for $\Psi$ to depend also on $g_{\mu\nu}$ leaves room for $\Psi^A$ to contain spinor fields if desired.\footnote{The same can be said of possible dependence of matter $u$'s transformations on the metric if $u$ contains spinor fields.  The 
 nonlinear group realization formalism avoids the  surplus structure of an orthonormal tetrad ($10$ components) and an extra local $O(1,3)$ gauge freedom ($6$ freedoms) to deprive the extra components of physical meaning ($16-6=10,$ the number of components of $g_{\mu\nu}$    
 \cite{OP,OPspinor,PittsSpinor,TAM2013TimeandFermionsConfProc}. The tetrad formalism is thus de-Ockhamized.   A common back door from the tetrad formalism to the nonlinear realization formalism is the imposition of the symmetric tetrad gauge condition, which is very popular in studying the Einstein-Dirac system of gravity plus electrons (or the like) and in supergravity \cite{vanNGauge,WoodardSymmetricTetrad} \cite[p. 234]{GatesGrisaruRocekSiegel}.   One could permit the $\Psi^A$ transformation rule to depend on matter fields $u$ as well, though it is hard to see why one would.}

The terms that survive when the gravitational and   material field equations are imposed takes the form 
\begin{eqnarray}  \int d^4x      \frac{ \delta S_m}{\delta \Psi^A} [\xi^{\mu} \partial_{\mu} \Psi^A - C^{A\nu}_{\mu}(\Psi, g?) \partial_{\nu} \xi^{\mu}] =0.
\end{eqnarray}
Integrating by parts, discarding boundary terms because the coordinate transformations are assumed to be trivial at the boundary, and using the arbitrariness of $\xi^{\mu}$ in the interior to remove the integration implies 
\begin{eqnarray} 
\frac{ \delta S}{\delta \Psi^A}  \partial_{\mu} \Psi^A   + \partial_{\nu} \left[ C^{A\nu}_{\mu}(\Psi, g?) \frac{ \delta S}{\delta \Psi^A} \right]=0.
\end{eqnarray}

Now let us make the simplest assumption, that the mental influence is described by a scalar field (just one function, the same in all coordinate systems;  a pseudoscalar, changing sign under negative-determinant coordinate transformations, would work the same way). In this case $C^{A\nu}_{\mu}(\Psi, g?)  =0.$  
One can consider various possibilities for $ \frac{ \delta S}{\delta \Psi}$; with only one component, $\Psi$ does not need the index $A$. 
Then the Lie derivative $\pounds_{\xi} \Psi$ along the vector field $\xi^{\mu}$ is just the directional derivative. 
The equation above becomes
\begin{eqnarray} 
\frac{ \delta S}{\delta \Psi} \xi^{\mu} \partial_{\mu} \Psi  =0.
\end{eqnarray} 
Because   $\xi^{\mu}$ is arbitrary due to general covariance, one therefore has 
\begin{eqnarray} 
\frac{ \delta S}{\delta \Psi} \partial_{\mu} \Psi  =0.
\end{eqnarray}
If  $ \frac{ \delta S}{\delta \Psi} \equiv 0$ (because $\Psi$ appears only in a boundary term), then $\Psi$ plays no role and can be eliminated from the discussion. Another possibility (which appears in a toy example below) is that $ \frac{ \delta S}{\delta \Psi} \equiv constant\neq 0$; then spatiotemporal constancy of $\Psi$ follows immediately.   
 Assuming then that $ \frac{ \delta S}{\delta \Psi}$ is not identically a constant ($0$ or not), suppose that $ \frac{ \delta S}{\delta \Psi}$ does not depend on $\Psi$ but depends on $g_{\mu\nu}$ and/or $u.$  Then wherever $\Psi$ varies spatiotemporally ($ \partial_{\mu} \Psi \neq 0$), as it must to have any chance of representing the mind's action on the body, the resulting equation  $ \frac{ \delta S}{\delta \Psi} (g_{\mu\nu}, u, \cancel{\Psi})=0$ will impose a novel restriction on the allowed states of matter and gravity like a Lagrange multiplier even without the \emph{postulation} that $\frac{ \delta S}{\delta \Psi}=0$.  (A nonzero constant $\Psi$ everywhere and always is not suitable for representing the influence of spatiotemporally localized minds, so if $\partial_{\mu} \Psi=0,$ the  boundary condition $\Psi=0$ implies that $\Psi$ vanishes throughout space-time.)
 Hence  $\Psi$ either does nothing or unreasonably restricts the physical possibilities for gravity and matter, or perhaps does nothing in some parts of space-time and unreasonably restricts the physical possibilities for gravity and matter  elsewhere; clearly this option fails.  Another possibility, which one might write loosely as $\frac{ \delta S}{\delta \Psi}(g_{\mu\nu}, u, \Psi \cancel{\partial_{\mu} \Psi}),$ is that  $\frac{ \delta S}{\delta \Psi}$ depends on $\Psi$ algebraically but not on derivatives of $\Psi.$  (In this case $\Psi$ is almost an auxiliary field, where an auxiliary field has Euler-Lagrange equations in which it appears algebraically and so can be solved for \cite{PonsSubstituting}.) In order that $\Psi$ not interfere with the dynamics of gravity and matter where and when $\Psi$ is `turned off' (such as outside brains), the  algebraic-in-$\Psi$ equation $\frac{ \delta S}{\delta \Psi}(g_{\mu\nu}, u, \Psi \cancel{\partial\Psi}) = 0$ must have as a solution $\Psi=0.$  While there might be other solutions as well---a cubic equation might have $3$ real solutions, \emph{etc.}---presumably they will be countable and will not fill an interval including  $\Psi=0.$  It seems impossible that $\Psi$ jump from $0$ to some nonzero solution of $\frac{ \delta S}{\delta \Psi}(g_{\mu\nu}, u, \Psi \cancel{\partial\Psi}) = 0$.  If $\Psi$ cannot slide continuously from $0$ to small nonzero values while satisfying the relevant algebraic equation, then this algebraic-in-$\Psi$ option does not work, either.   
The remaining possibility (at least if non-local coupling is excluded) is that $\frac{ \delta S}{\delta \Psi}(g_{\mu\nu}, u, \Psi ) = 0$  depends on $\partial_{\mu}\Psi;$ it might or might not depend algebraically on $\Psi$ as well.

An example might help:  let $\Psi$ appear in the Lagrangian density \emph{via} the expression $-\frac{1}{2} \sqrt{-g} g^{\mu\nu} (\partial_{\mu}\Psi) \partial_{\nu} \Psi,$ like a massless scalar field.  Then the equation $\frac{ \delta S}{\delta \Psi} \partial_{\mu} \Psi  =0$ implies that at any space-time point either $\Psi$ is spatio-temporally constant or it satisfies the wave equation $\partial_{\mu} (\sqrt{-g} g^{\mu\nu} \partial_{\nu} \Psi) =0.$  While most solutions of the wave equation do vary spatiotemporally (a step in the right direction), they are also fixed by initial and boundary data.  Prior to the existence of minds acting on bodies,  $\Psi$ and  its time derivative should vanish, yielding $\Psi=0$ everywhere and always due to the wave equation.   Hence this partial differential equation is still too strong to leave $\Psi$ the wiggle room needed to represent the soul's action on the body.   Assuming that the wave equation example is representative, it seems that local classical field theory provides no options suitable for modelling the effect of the mind on the body if such effects are mediated by a single scalar (or pseudoscalar) function.\footnote{One might consider the relevance of solitons (as an anonymous reader has suggested).  It would seem that solitons \cite{SolitonsManton} have little relevance to this argument, because most types (except Skyrmions) either are not relevant to situations varying nontrivially in three spatial and one time dimension or involve so many fields involved that the Bianchi identities would be saturated so that there is no difficulty for the mind to act on matter.  Moreover, most or all solitons are only approximately localized spatially and temporally.  Merely approximate localization would imply that my soul could act on matter before Julius Caesar crossed the Rubicon (but only very weakly)  and in the Andromeda Galaxy (but only very weakly), which seems absurd and which might make telekinesis and similar phenomena  common.  }
 Hence either there is no such dualist mental causation after all, or its influence is not represented by one scalar (or pseudoscalar) field, or some other Lagrangian makes a relevant qualitative difference.  That result contrasts sharply with what would happen in a special relativistic field theory, namely, that the energy-momentum currents would fail to be conserved where and when the mental causation occurs. Regardless of one's interpretation of gravitational energy and general relativistic conservation laws, General Relativity makes mental causation more difficult by excluding at least the simplest case.

What if there are two, three or four scalar fields?  If their gradients are linearly independent, then one will have two, three or four copies (respectively) of the same sad story.  One can use the scalars with independent gradients as space-time coordinates locally and thus turn the $\partial_{\mu} \Psi^A$ expressions into a rectangular matrix with two, three or four entries with the value $1$ and the rest $0$.  But if the scalars' gradients are not all linearly independent, then one has correspondingly fewer equations and correspondingly fewer unknowns.  So evidently even up to four scalar fields cannot do the job; success, if possible at all, would require more scalars (at least $5$), non-scalars, or some combination of the two.

Withdrawing the simple assumption that $\Psi$ is a scalar field and allowing it to be a quite arbitrary collection of fields $\Psi^A,$ one has the more general relation   
   \begin{eqnarray*} 
  \frac{ \delta S}{\delta \Psi^A}  \partial_{\mu} \Psi^A   + \partial_{\nu} \left[ C^{A\nu}_{\mu}(\Psi, g?) \frac{ \delta S}{\delta \Psi^A} \right]=0.
   \end{eqnarray*}
from above.  
The question arises whether these relations are still strong enough to annihilate the soul's influence.  Note that there are \emph{still only four of these identities},\footnote{One could find a fifth identity by considering electromagnetic gauge freedom and maybe a few more (perhaps irrelevant ones \cite{PittsArtificial,FrancoisArtificialSubstantialGauge}) from the weak and strong nuclear forces. While the details might be changed, the conceptual point is not affected.} but there is \emph{no limit} on how many field components over which the index $A$ can range.  The freedom to choose the form of the transformations of the fields is also relevant, albeit in a less transparent and less important way.  In the single scalar $\Psi$ case, the generalized Bianchi identities exclude  the soul's influence, which is forced to constancy and then vanishing everywhere. But if $\Psi^A$ has enough components ($5$ or perhaps a bit higher being a plausible first estimate), then arguably the soul `wins' and manages to act as it wishes even in a general relativistic world.  Perhaps the soul deploys four scalar fields sacrificially  as pawns to saturate the generalized Bianchi identities while the more commanding field component(s) undertake to implement the soul's intentions in the world. 
If there are, for example, $5$ scalar fields in $\Psi^A,$ let $\Psi^4$ try to impose the soul's intentions on the world (or at any rate on the brain).  Can it get by with a little help from its friends, the other four scalar fields?  It seems plausible that one (call it $\Psi^0$) could accommodate dependence on coordinate $x^0,$ another (call it $\Psi^1$) could accommodate dependence on $x^1,$ another (call it $\Psi^2$) on $x^2,$ and another (call it $\Psi^3$) on $x^3,$ locally, for example, so as to satisfy the generalized Bianchi identities without forcing  $ \frac{ \delta S}{\delta \Psi^4}  \partial_{\mu} \Psi^4 
=0$.  That last equation  would seem to block the soul's effort. (This would not be the first time in General Relativity that one needs to invest $5$ scalar fields to keep $1$ of them \cite[pp. 253, 254]{BergmannHandbuch} \cite[pp. 81, 82]{RovelliBook}.)
This function-counting argument is admittedly crude and merely plausible.  A more definitive resolution would require a more specific postulate about how many field components of various types $\Psi^A$ contains and how they couple to the metric $g_{\mu\nu}$ and matter fields $u$.  Thus it could turn out that the general relativistic objection still has force even with some additional field components to employ.  But that seems unlikely to be true to any great extent.  There is a literature on underdetermined differential equations, which in general outline confirms the expectation that having fewer equations than unknowns leads to having a variety of solutions  \cite{BenderUnderdeterminedPDE,ElkinUnderdeterminedODE}.

\emph{Mutatis mutandis} a similar discussion would apply to (special) divine action (a.k.a. miracles),  angelic action,  demonic action,  genie action (genies/jinn being part of Muslim belief),  action by the spirits of the dead (which was accepted by a number of distinguished Victorian scientists who studied the matter in s\'{e}ances), or action of whatever wholly or partly immaterial personal beings someone might wish to consider.  Note that the description above does \emph{not} supply a field $\Psi$ for \emph{each} spirit.  If there can be multiple angels in the same place (a topic of renewed interest recently in analytic philosophy \cite{HawthorneAngelsNeedle}), then one could consider how to describe such scenarios.  But the discussion has strayed near, if not past, the boundaries of contemporary plausibility structures already.

Some readers  might share Carroll's experience:  ``[n]obody ever asks these questions out loud, possibly because of how silly they sound.'' While I don't altogether disagree, in my estimation the asking and answering of these questions pays for itself, if in no other way, by the fruit yielded regarding gravitational energy in General Relativity and the consequences for conserved quantity theories of causation.  One might also respond that the questions have a sort of timeless interest; a few centuries  centuries ago such questions, or at any rate close analogs of such questions, were very seriously  debated by leading thinkers  \cite{WatkinsPhysicalInfluxCrusius,WatkinsHarmonyLeibnizPhysicalInflux}, whereas some current questions (such as how consciousness is possible for a purely physical system) not only sounded silly to many people, but attracted strong arguments in refutation, such as Leibniz's mill argument:  
\begin{quote} 
Furthermore, one is obliged to admit that \emph{perception} and what depends upon it is \emph{inexplicable on mechanical principles}, that is, by figure and motions.  In imagining that there is a machine whose construction would enable it to think, to sense, and to have perception, one could conceive it enlarged while retaining the same proportions, so that one could enter into it, just like into a windmill.  Supposing this, one should, when visiting within it, find only parts pushing one another, and never anything by which to explain a perception.  Thus it is in the simple substance, an not in the composite or in the machine, that one must look for perception. \cite[p. 83]{LeibnizMonadMachine} \end{quote} 
This kind of argument has contemporary proponents \cite{PlantingaMaterialism,RuseLeibnizMachineReligionNotNewAtheism}. 
Without endorsing the argument, Markus Gabriel  calls it ``one of the earliest modern formulations of the so-called {\bf hard problem of consciousness}'' \cite[p. 121, emphasis in the original]{GabrielNotaBrain}. 
The de-materialization of matter into fields with basically just mathematical properties seems not to affect the force of the argument.  
 Whether one struggles to take seriously a mathematical description of the mind-body problem in General Relativity due to materialist sympathies or not, it is  sometimes useful to step out of one's view and see to what extent the weight of argument forces one back into it. 
The analogy of destructive testing, understanding something by breaking it, again comes to mind.  
 One non-dualist undertook such an exercise and concluded that the arguments against dualism weren't very compelling \cite{LycanDualism}.  Here I have taken Carroll's foundling research questions (or analogs suited to classical field theory) seriously enough to do some calculations and find some results that, in some cases, are very congenial to Carroll's views.


\subsection{More Options from Inequivalence under Field Redefinitions}

Another surprise is that there are  more \emph{distinct} ways that the distinctively mental influence $\Psi$ on matter might be realized than one might have expected, because some options that are equivalent in more standard physical contexts become distinct in the present problem.  Suppose that one is trying to write down a toy mathematical description of how the distinctively mental influence (the soul's effect) relates to gravity and matter.  The interaction term coupling the soul to gravity and matter should respect general covariance and thus be a scalar under coordinate transformations.  Hence the integrand should be a scalar density of weight $1$. Ignoring matter and letting the soul couple to gravity only for simplicity (to make an important point vivid), two obvious choices (out of an uncountable infinity) are: 
$$ \int d^4x \sqrt{-g}(x) \Psi(x) $$ 
(the soul's influence $\Psi$ acts as a scalar field, the same in all coordinate systems), or 
$$ \int d^4x \tilde{\Psi}(x)$$ 
(the soul's influence acts as a scalar density of weight $1$).  Usually in physics one regards such choices as equivalent because one can relate the two choices by $\tilde{\Psi} = \sqrt{-g} \Psi.$  In cases where all fields satisfy the principle of least action, it is easy to find that the resulting equations of motion are unaffected by field redefinitions. An analogous result holds in quantum field theory \cite{DuffQFTinCSTinconsistent}.  But now suppose (as holds in our case) that some fields satisfy the principle of least action and some do not.
  A typical application is electromagnetism in special relativity with arbitrary coordinates; the flat metric tensor of Special Relativity appears in a mildly non-trivial form because one can label the flat space-time in an arbitrary way (coordinates can be angles or can slosh around because your body or the Earth or the Sun defines the origin, or whatever).  
A variation of the action $S$ made by varying the fields $A_{\mu}$ and $\eta^{\mu\nu}$ (the inverse of $\eta_{\mu\nu}$), with the variations being $0$ at the boundary, is 
\begin{eqnarray} 
\delta S= \int d^4x \left( \frac{\delta S}{\delta A_{\mu}}(x) \cdot \delta A_{\mu}(x)  + \frac{\delta S}{\delta \eta^{\mu\nu} }|A \cdot \delta \eta^{\mu\nu}(x) \right), 
\end{eqnarray}
where $|A$ indicates the holding constant of $A_{\mu}.$  The Euler-Lagrange equations for electromagnetism are $ \frac{\delta S}{\delta A_{\mu}}(x)=0.$  Now let us introduce  a new field $B^{\nu} = A_{\mu} \eta^{\mu\nu}.$  
One can use $B^{\nu}$ and $\eta^{\mu\nu}$ as a new set of fields (so that $B^{\nu}$ is now primitive and the relation  $B^{\nu} = A_{\mu} \eta^{\mu\nu}$ now defines  $ A_{\mu}$) and vary the action:  
\begin{eqnarray} 
\delta S= \int d^4x \left( \frac{\delta S}{\delta B^{\nu}}(x) \cdot \delta B^{\nu}(x)  + \frac{\delta S}{\delta \eta^{\mu\nu} }|B^{\nu} \cdot \delta \eta^{\mu\nu}(x) \right), 
\end{eqnarray}
with Euler-Lagrange equations  $  \frac{\delta S}{\delta B^{\nu}}(x)=0$.
The relation $B^{\nu} = A_{\mu} \eta^{\mu\nu}$ lets one write $\delta B^{\nu} = A_{\mu} \delta \eta^{\mu\nu} +  \eta^{\mu\nu} \delta A_{\mu}.$  Equating coefficients of $A_{\mu}$ and of $\delta \eta^{\mu\nu}$ yields, respectively, 
\begin{eqnarray} 
\frac{ \delta S}{\delta A_{\mu} } = \frac{\delta S}{\delta B^{\nu} } \eta^{\mu\nu},  \\
\frac{\delta S}{\delta \eta^{\mu\nu} }|A =  \frac{\delta S}{\delta \eta^{\mu\nu} }|B  + \frac{\delta S}{\delta B^{(\nu} } A_{\mu)}
\end{eqnarray} 
(where the parentheses imply symmetrization).  
One finds that  the electromagnetic field equations are equivalent, and the (Hilbert-Rosenfeld) stress-energy tensor (basically $\frac{\delta S}{\delta \eta^{\mu\nu}}$) changes only by terms that are $0$ when the electromagnetic field equations hold \cite{PonsEnergy,MassiveGravity3}. 
In this case one has redefined the dynamical field by folding  some non-dynamical field into it ($B^{\nu} = \eta^{\mu\nu} A_{\mu}$), but the non-dynamical inverse flat metric tensor $\eta^{\mu\nu}$ is left alone. Hence it doesn't matter if electromagnetism is a covariant vector $A_{\mu}$ or a contravariant vector $B^{\nu}.$  The usual covariant choice has the virtue of not requiring the construction of covariant derivatives and so perhaps is preferable.

But the case of the soul's influence on gravity (and other physical fields) is different, because the entity not redefined, gravity, has Euler-Lagrange equations but the redefined entity does not. By an analogous derivation one finds that new and old Euler-Lagrange equations differ.   Does the soul's influence transform as (\emph{e.g.}) a scalar or a scalar density of weight $1$? (It should  be possible to describe the soul's influence in any admissible coordinate system within General Relativity, so questions of this sort have to make sense, however incongruous they might sound at first.)   Now the dynamical field $\sqrt{-g}$ (which defines volumes in General Relativity), for which the principle of least action applies (because it is part of the space-time metric $g_{\mu\nu}$), is left alone, while the mental influence $\Psi,$ which does \emph{not} satisfy the principle of least action, is redefined by folding in some of $\sqrt{-g}.$ Using the generalized Bianchi identities as above, one can easily find that the term $ \int d^4x \sqrt{-g}(x) \Psi(x) $ (the soul acting as a scalar) leads to a cosmological constant ($\Psi=constant$), altering the gravitational field equations in the same way throughout the whole history of the universe, which is startling for an influence that was supposed to be confined to my brain (and my lifetime)---unless one imposes the boundary condition $\Psi=0$ at the boundaries, in which case $\Psi=0$ everywhere and always, so the soul does nothing at all.  The term $ \int d^4x \tilde{\Psi}(x)$ (the soul's influence as a scalar density of weight $1$), by contrast, definitely does nothing at all (even without appeal to boundary conditions) because it does not couple to anything physical; that is also disappointing, but  in a \emph{different} way. Not appealing to boundary conditions in one case and needing boundary conditions in another shows that the two cases are inequivalent.

 The point is not that two ludicrously  oversimplified examples behave badly, but that they are \emph{inequivalent}, one not needing boundary conditions and the other needing them---contrary to what one might have assumed based on related but subtly different experiences with field redefinitions in fundamental physics.  
The two cases' Euler-Lagrange equations for gravity (the Einstein-like equations) differ.  Explicitly one has for the variational derivatives
\begin{eqnarray}
  0 \stackrel{?}{=}  \frac{\delta S}{\delta \sqrt{-g} }|\Psi =  \frac{\delta S}{\delta \sqrt{-g} }|\tilde{\Psi}   +\frac{\delta S}{\delta \tilde{\Psi} } \frac{\tilde{\Psi}}{\sqrt{-g}} \stackrel{?}{=} 0 ,  \\
\frac{\delta S}{\delta \Psi} = \frac{\delta S}{\delta \tilde{\Psi}} \sqrt{-g} \neq 0.
\end{eqnarray}
The first equation, which shows $10\%$ of Einstein's equations (the trace) in the presence of mental causation, shows that the gravitational field equations differ by a term involving the soul's influence   and thus are not equivalent.  Whether the soul's influence is a scalar, or a scalar density of weight $1$, or something(s) else,   makes a difference in the field equations. 
One might have expected $ \frac{\delta S}{\delta \sqrt{-g} }|\Psi = 0$ to be equivalent to $  \frac{\delta S}{\delta \sqrt{-g} }|\tilde{\Psi} =0,$ but one sees that in fact they differ by a nonzero term $ \frac{\delta S}{\delta \tilde{\Psi} } \frac{\tilde{\Psi}}{\sqrt{-g}}.$ 
Admittedly, the generalized Bianchi identities substantially reduce this difference by forcing the soul's influence (rescaled by a suitable power of $\sqrt{-g}$) to be constant, but do not go so far as to establish equivalence of the field equations, which differ by a cosmological constant.  (Imposing boundary conditions to make that constant influence $0$ does yield equivalence, but that is too weak:  using boundary conditions is inequivalent to not using them.)   Analogously, one needs to contemplate in principle distinct vector-like possibilities:  a covariant vector, a contravariant vector,  a covariant vector density of any real weight, and a contravariant vector density of any real weight are not one option, but $2 + 2\infty$ that might differ.  Obviously there are other possibilities: various flavors of spinor fields, \emph{etc.}, not to mention multiple scalars, multiple scalar densities, a scalar and vector of some sort, \emph{etc.} Thus not only can $\Psi^A$ involve any kinds of fields in whatever quantities you like, but also some cases that one might have expected to be equivalent, in fact differ.  Hence there is in principle an enormous zoo of possibilities that could be explored if one aimed (as I do not) to try to give some positive realistic account of how immaterial minds might act in a general relativistic physical world.    If one wishes to supply Carroll's missing argument, then one has a great many distinct options to consider. 
 If some options survives this elimination process, the result will still be very far from biological realism, presumably.  But that is a different kind of argument, one best left to neuroscientists rather than theoretical physicists and philosophers of physics.


\section{Whom Does the General Relativistic Objection Affect?}

The new general relativistic objection from the Bianchi identities (perhaps one can say, from energy conservation) is not  obviously question-begging given Noether's first theorem and its converse.
Thus it is a better argument than the traditional Leibnizian one.  But how  good an objection is it?  Against which kinds of interactionist substance dualists (or non-epiphenomenalist property dualists) is it effective?  

As noted above, the power of the Bianchi identities to constraint the mental influence $\Psi^A$ depends especially on the number and to some extent on the type of fields included in $\Psi^A$. General Relativity implies (in $4$ space-time dimensions) 4 Bianchi identities. General Relativity thus puts up a nonzero but bounded amount of resistance to external forces, unlike earlier theories that put up no resistance at all.   But what determines how many components there are in $\Psi$?  The answer, surely, depends on what kind of process produced human beings.

 If it was a purely \emph{naturalistic} evolutionary process, then any  immaterial self must evolve naturally.
Such is, one might think, impossible.  If so, then dualism is already ruled out without the need to appeal to the Bianchi identities to constrain interactionism.  If perchance an immaterial self could evolve, perhaps evolving components of $\Psi^A$ is difficult and hence improbable; maybe the physicists' exponentially suppressed probability from statistical mechanics is relevant? Then $\Psi^A$ should have few components; perhaps $5$, roughly the minimum number needed to overwhelm the Bianchi identities, is too many? (It is difficult to speak sensibly  about this question; perhaps one should remain silent.)  Or if it is both possible and not difficult to evolve $\Psi^A,$ it will still be random (apart from natural selection, which is not the productive phase of evolution), not engineered so as to give  $\Psi^A$ enough components to overwhelm the Bianchi identities and let the soul act on the world.  So it would seem that if souls somehow manage to evolve naturally and acquire field components $\Psi^A$ with which to try to act on the physical world, then there is still a reasonable chance that the Bianchi identities will drown $\Psi^A$,  forcing it to be $0$.  The prospects for interactionism given a purely naturalistic evolutionary process  do not seem all that bright.

And yet such a conclusion seems to conflict with the views of some interesting people.  If one attempts to look for nontheists (for whom \emph{naturalistic} evolution will be required) who defended or at least sympathized somewhat with substance dualism or allied doctrines about spirits, one does not come up empty in  the  20th century.  Idealist  McTaggart \cite{McTaggartImmaterialismNathan} is perhaps appropriately listed, though whether his spirits acted on matter would require care.   Broad was doubtful about a personal God \cite{BroadPsychical}, sympathetic to parapsychology \cite[p. 395]{Broad} \cite{BroadPsychical}, and  unwilling to criticize substance dualism on grounds of energy conservation  \cite[pp. 106-108]{Broad} (though his response to that argument was fallacious \cite{EnergyMental}). Turing took the statistical evidence for telepathy to be overwhelming \cite{TuringESP}.  More recent figures include  Popper  \cite{PopperEccles} and parapsychology expert John Beloff  \cite{BeloffExistenceMind,SmythiesBeloff}.  Beloff  intended  to be surprised if he still existed after death (though not annoyed like Broad!) \cite[p. 13]{SteinkampBeloff} and was an atheist \cite{BeloffObituaryBraude,BeloffObituaryBlackmore}.  Some more recent authors also aim to  incorporate spiritual/parapsychological phenomena or entities within a broadened naturalist rather than  supernaturalist framework \cite{GriffinParapsychology,LevitationGrossoCopertino}.
Against people who affirmed causally active spirits but not God, the general relativistic objection may have some force.  This category should perhaps also contain people influenced by nontheistic Buddhism.  Perhaps at least some  who  work on evidence for  reincarnation  (a topic on which Carl Sagan had respectful words  \cite[pp. 47, 48]{SaganBrocaStevenson}) (\emph{e.g.}, \cite{CasesReincarnationType,StevensonReincarnationBiologyBirth,PasrichaEmpirical,PasrichaEvidence}) would also qualify, though I cannot comment on Stevenson's or Pasricha's views.  Thus one should not exaggerate the connection between dualism and  theistic belief, because parapsychologists who take themselves to be doing science frequently are not theists.

Because the strength of the Bianchi identity-based objection to Cartesian mental causation diminishes, the more fields are at the soul's disposal, it follows that the seriousness of the objection winds up being interestingly related to the theism-atheism debate.  Without God, a soul's acquiring fields of influence on the physical world is presumably impossible, or difficult, or random, and hence has a reasonable chance of not overcoming the tendency  of the Bianchi identities to trivialize the mental influence. But given theism of a traditional stripe (perhaps a form of theistic evolution), a soul's possessing such fields appears to be possible,  not difficult, and  designed for the task at hand, hence highly likely to succeed in overcoming the Bianchi identities to let the soul act on the physical world.   Perhaps the import of the new general relativistic objection, then,  is that \emph{atheists shouldn't be interactionist dualists}.  Of course not many atheists have been interactionist dualists anyway, at least not in the West, due to arguments not involving conservation laws.   There are prominent examples from the recent past, however, as noted above.  
 The new result from General Relativity might put pressure on views that affirm souls but do not affirm God (such as some forms of Buddhism and perhaps forms of spiritualism) or that claim to incorporate spiritual/parapsychological phenomena or entities within a broadened naturalist rather than  supernaturalist framework (\emph{e.g.}, \cite{BroadPsychical,GriffinParapsychology,ParapsychologyHandbook21st,LevitationGrossoCopertino}).  But at least for  theistic views (whether Abrahamic, Hindu or some other kind) and naturalism, the new general relativistic objection does not ultimately make much difference for most people: typically either the objection doesn't work, or its conclusion was already accepted anyway on other (comparably good) grounds.  In that sense, the new general relativistic energy conservation objection  and the old Leibnizian energy conservation objection  are somewhat alike:  neither works very well in motivating many people to change their beliefs, though the general relativistic objection is better because it does discourage some views, namely those involving causally active souls without God.

On the other hand, it is noted by people of various views  that there is a certain natural harmony between dualism and theism  \cite[chapter 7]{BainMind} \cite{TaliaferroConsciousnessGod,McGinnMysteriousFlame,FosterSoulBody,WiebeGodSpirits,PlantingaMaterialism} \cite[p. 352]{MorelandRaeBodySoul}  \cite[pp. 262-265]{MorelandCraigWorldview}.  Thus objections to interactionism that presuppose nontheism will not be all that effective against this large class of  substance dualists. Supposing that God exists and (\emph{e.g.}) somehow guided evolution, presumably there are  enough of and the right kinds of $\Psi^A$ components to overcome the restriction imposed by the Bianchi identities. 
Perhaps there is an imperfect analogy to this remark:  ``Nature may shout no, but human ingenuity\ldots may always be able to shout louder.'' \cite{LakatosHistory} 
 Some examples from the 18th century German debate on Leibniz's objection and pre-established harmony are useful.  
Johann Peter Reusch, a Wolffian, rejected  pre-established harmony partly  on the grounds that it made bodies superfluous. 
He concluded that there is no sufficient reason for creating bodies \cite{WatkinsHarmonyLeibnizPhysicalInflux}, a serious objection for a Wolffian.   
Likewise  Christian August Crusius rejected the conservation laws, \begin{quote} noting that if they were true, the absurd results would follow that minds could not cause any motion and, as Reusch had noted earlier, that matter would not be able to fulfill the purpose for which God intended them [\emph{sic}], namely to be a means for rational and free beings. \cite{WatkinsHarmonyLeibnizPhysicalInflux}. \end{quote}   The attached footnote translates Crusius:  
``But then the material world would be of no use to minds, and [it] would have been created completely without a purpose'' \cite[footnote 164]{WatkinsHarmonyLeibnizPhysicalInflux} \cite[footnote 98]{WatkinsPhysicalInfluxCrusius}. 
To be sure, nowadays there exist Christians who take physicalist views about human persons  \cite{PVIDualismMaterialism,RudderBakerDualism,MurphySoul,MerricksResurrection,Corcoran} and so could evade Crusius's argument.  %

Perhaps the import of the development of Carroll's  foundling is that  \emph{nontheists} have an additional reason to deny causally efficacious souls. This point isn't entirely uninteresting  because it undermines a class of views that has in fact been and still is held in some circles.  (That result contrasts with Leibniz's objection, which fails entirely even at the classical level, making appeal to quantum mechanics unnecessary \cite{EnergyMental}.)      
  In any case  General Relativity does not make mental causation easier, but can make it harder in some contexts.   Nontheistic dualists therefore might need to hope for the truth of the sometimes-heard claim that  quantum mechanics  facilitates soul-to-body causation.


\section{Philosophical Payoff Outside the Philosophy of Mind} 

While one might think that the relevance of relating General Relativity to the traditional mind-body problem and especially the Leibnizian energy conservation strand would not propagate beyond the philosophy of mind and associated sectors of metaphysics, it appears that there is in fact a payoff in two other areas, one in the foundations of physics and one in philosophical theories of causation, as hinted earlier. 


\subsection{Bearing on Gravitational Energy (Anti)Realism and Conservation}

The treatment of the generalized Bianchi identities above shows that the relevance of General Relativity for mental causation does not have to be yoked to the frequently denigrated pseudotensor conservation laws for energy  and momentum.  Rather, the tendency of General Relativity to restrict immaterial-to-material causation can be explored in a tensorial way, free of conventional taint and hence free of the usual interpretive controversy over gravitational energy.  The fact that these calculations do show a tendency to resist immaterial-to-material causation shows that the heuristic force of the widely received anti-realist gloss on the conservation laws is incorrect.  That heuristic was the basis for the Mohrhoff-Collins view that General Relativity facilitates immaterial-to-material causation.  The conclusion that General Relativity makes immaterial-to-material causation harder rather than easier could have been motivated (and indeed has been \cite{EnergyGravity}) by realism about gravitational energy localization and consequently about total (gravitational + material) energy-momentum conservation. Anti-realism continues to attract philosophical adherents \cite{DuerrFantasticBeasts}.  But because anti-realism  leads to a false heuristic and realism leads to a correct heuristic in this context, there results some additional justification for taking realism about gravitational energy and conservation laws more seriously.  General Relativity \emph{loves} conservation of energy-momentum; indeed it isn't a large exaggeration to say that General Relativity \emph{is} energy-momentum conservation, given that the field equations and the conservation laws are logically equivalent.  (This outcome charmingly satisfies one of Einstein's desiderata \cite{BradingConserve,EinsteinEnergyStability}.  There are presumably other field equations than Einstein's, built with higher derivatives of the metric, with the same property but a different form of the conservation laws, so the claim is something of an exaggeration.  Such theories will share some important structural properties with Einstein's theory, however.)  One can interpret the general relativistic objection to immaterial-to-material causation as saying that General Relativity tries hard to conserve energy and momentum (infinitely many of each, at each point in space!), much harder than other field theories do.  Thus contemplation of spirit-to-matter causation sheds light on the gravitational energy localization debate in favor of realism. 


\subsection{Bearing on Conserved Quantity Theories of Causation }

As various authors have noted, it is rather awkward for conserved quantity theories of causation that energy and momentum are no longer conserved quantities given the widespread anti-realism about conservation laws in light of General Relativity \cite{RuegerCauseEnergy,CurielCauseGR,LamCauseConservedGR}.   
Dowe has responded to this concern of Rueger's as follows: 
\begin{quote} 
But that there are general relativistic spacetimes in which global conservation laws do not hold does not entail that global conservation laws fail in our world. Whether they do or not depends on the actual structure of spacetime, and in particular whether certain symmetries hold. As I understand it, our spacetime does exhibit the right symmetry, and that [\emph{sic}] global conservation laws do hold in our universe as far as we know. I take it, then, that the conserved quantity theory is not refuted.

 I suggested that the account holds in all physically possible worlds, that is, in all worlds which have the same laws of nature as ours. Has Rueger shown that this is not so? Not at all. To say, for example, that non-symmetric spacetimes are possible can be misleading. It means simply that it is a solution to the equations of the General Theory of Relativity. But this doesn't mean that such a world is a physically possible world in the sense given above. If such a world violates other laws that hold in the actual world, then that world is not physically possible. This is exactly what we have in these non-symmetric spacetimes. Symmetries and conservation laws that hold in the actual world break down, so it is not a physically possible world in my sense. 

Therefore we need not give up on the Conserved Quantity theory, understood as a contingent hypothesis. \cite{DoweDefended} 
\end{quote} 
 Dowe seems not to realize that maintaining such a global conservation law would be revisionary---indeed he is suggesting in effect that the usual cosmological models are not even physically possible---so the cost is greater than he envisaged.  Whether a global conservation law is philosophically interesting is difficult to judge.  Many philosophers are unaware that global conservation laws are a crude and archaic relative of the conservation laws that are of primary interest in physics, and hence are too easily satisfied with global conservation laws.

If the above considerations have shown that realism about gravitational energy localization and consequently about local conservation laws in General Relativity has something going for it, then that is a boon for a conserved quantity theory of causation.  If energy and momentum are no less conserved in General Relativity than in other field theories, then Rueger's objection no longer holds.  One only needs to become accustomed to speaking of plural energies and momenta.


\section{Acknowledgments}

This work is supported by the John Templeton Foundation, grant \#60745.  It does not necessarily reflect the views of the  Foundation.  I also thank  Tim Crane, Jeremy Butterfield, Sam Newlands,  Robert Garcia, Alex Arnold, Marcin Iwanicke, Ole Koksvik, Matthew Lee, William Simpson, and Chris Meyns for helpful references  and Alexandre Guay for comments.  All views are my own.


\begin{thebibliography}{}

\bibitem[Anderson, 1958]{AndersonPrimary}
Anderson, J.~L. (1958).
\newblock Reduction of primary constraints in generally covariant field
  theories.
\newblock {\em Physical Review}, 111:965--966.

\bibitem[Anderson, 1967]{Anderson}
Anderson, J.~L. (1967).
\newblock {\em Principles of Relativity Physics}.
\newblock Academic, New York.

\bibitem[Atmanspacher, 2015]{Atmanspacher}
Atmanspacher, H. (2015).
\newblock Quantum approaches to consciousness.
\newblock In Zalta, E.~N., editor, {\em Stanford Encyclopedia of Philosophy}.
\newblock
  https://plato.stanford.edu/archives/sum2015/entries/qt-consciousness/.

\bibitem[Averill and Keating, 1981]{AverillKeating}
Averill, E. and Keating, B.~F. (1981).
\newblock Does interactionism violate a law of classical physics?
\newblock {\em Mind: A Quarterly Review of Philosophy}, 90:102--107.

\bibitem[Bain, 1873]{BainMind}
Bain, A. (1873).
\newblock {\em Mind and Body: The Theories of Their Relation}.
\newblock Henry S. King \& Co., London.

\bibitem[Baker, 1995]{RudderBakerDualism}
Baker, L.~R. (1995).
\newblock Need a {Christian} be a mind/body dualist?
\newblock {\em Faith and Philosophy}, 12:489--504.

\bibitem[Bastianelli and {van Nieuwenhuizen}, 2006]{vanNPathCurved}
Bastianelli, F. and {van Nieuwenhuizen}, P. (2006).
\newblock {\em Path Integrals and Anomalies in Curved Space}.
\newblock Cambridge University Press, Cambridge.

\bibitem[Bauer, 1918]{BauerEnergy}
Bauer, H. (1918).
\newblock \"{U}ber die {Energiekomponenten des Gravitationsfeldes}.
\newblock {\em Physikalische Zeitschrift}, 19:163--165.

\bibitem[Beloff, 1962]{BeloffExistenceMind}
Beloff, J. (1962).
\newblock {\em The Existence of Mind}.
\newblock MacGibbon and Kee, London.

\bibitem[Belot, 2006]{BelotConservation}
Belot, G. (2006).
\newblock Conservation principle.
\newblock In Borchert, D.~M., editor, {\em Encyclopedia of Philosophy}, volume
  2: Cabanis--Destutt de Tracy, pages 461--464. Thomson Gale, second edition.

\bibitem[Bender et~al., 2000]{BenderUnderdeterminedPDE}
Bender, C.~M., Dunne, G.~V., and Mead, L.~R. (2000).
\newblock Underdetermined systems of partial differential equations.
\newblock {\em Journal of Mathematical Physics}, 41:6388--6398.

\bibitem[Bergmann, 1958]{BergmannConservation}
Bergmann, P.~G. (1958).
\newblock Conservation laws in general relativity as the generators of
  coordinate transformations.
\newblock {\em Physical Review}, 112:287--289.

\bibitem[Bergmann, 1962]{BergmannHandbuch}
Bergmann, P.~G. (1962).
\newblock The general theory of relativity.
\newblock In Fl\"{u}gge, S., editor, {\em Prinzipien der Elektrodynamik und
  Relativit\"{a}tstheorie}, volume~IV of {\em Handbuch der Physik}, pages
  203--272. Springer, Berlin.

\bibitem[Birrell and Davies, 1982]{BirrellDavies}
Birrell, N.~D. and Davies, P. C.~W. (1982).
\newblock {\em Quantum Fields in Curved Space}.
\newblock Cambridge University Press, Cambridge.

\bibitem[Blackmore, 2006]{BeloffObituaryBlackmore}
Blackmore, S. (16 June 2006).
\newblock {John Beloff Obituary}.
\newblock {\em The Times}.
\newblock https://www.susanblackmore.uk/journalism/john-beloff-obituary/.

\bibitem[Bondi, 1957]{BondiNature}
Bondi, H. (1957).
\newblock Plane gravitational waves in general relativity.
\newblock {\em Nature}, 179:1072--1073.

\bibitem[Born, 1914]{BornEnergyMieHerglotz}
Born, M. (1914).
\newblock {Der Impuls-Energie-Satz in der Elektrodynamik von Gustav Mie}.
\newblock {\em Nachrichten von der K\"{o}niglichen Gesellschaft der
  Wissenschaften zu G\"{o}ttingen, Mathematisch-Physikalische Klasse}, pages
  23--36.
\newblock Translated as {``The Momentum-Energy Law in the Electrodynamics of
  Gustav Mie''} in {J\"{u}rgen Renn and Matthias Schemmel}, editors, \emph{{The
  Genesis of General Relativity, Volume 4: Gravitation in the Twilight of
  Classical Physics: The Promise of Mathematics}}, {Springer, Dordrecht (2007),
  pp. 745-756}.

\bibitem[Brading, 2001]{BradingDissertation}
Brading, K. (2001).
\newblock {\em Symmetries, Conservation Laws, and Noether's Variational
  Problem}.
\newblock PhD thesis, University of Oxford.

\bibitem[Brading, 2005]{BradingConserve}
Brading, K. (2005).
\newblock A note on general relativity, energy conservation, and {Noether's}
  theorems.
\newblock In Kox, A.~J. and Eisenstaedt, J., editors, {\em The Universe of
  General Relativity}, Einstein Studies, volume 11, pages 125--135.
  Birkh\"{a}user, Boston.

\bibitem[Braude, 2006]{BeloffObituaryBraude}
Braude, S. (4 July 2006).
\newblock {John Beloff: Scientist} who put parapsychology on the academic map.
\newblock {\em The Guardian}.
\newblock
  https://www.theguardian.com/science/2006/jul/04/obituaries.guardianobituarie%
s.

\bibitem[Braude, 1986]{BraudeLimitsInfluence}
Braude, S.~E. (1986).
\newblock {\em The Limits of Influence: Psychokinesis and the Philosophy of
  Science}.
\newblock Routledge \& Kegan Paul, New York.

\bibitem[Broad, 1937]{Broad}
Broad, C.~D. (1937).
\newblock {\em The Mind and its Place in Nature}.
\newblock Tarner Lectures, Trinity College, Cambridge, 1923. Kegan Paul,
  Trench, Trubner and Co., London.

\bibitem[Broad, 1953]{BroadPsychical}
Broad, C.~D. (1953).
\newblock {\em Religion, Philosophy and Psychical Research: Selected Essays}.
\newblock Routledge \& Kegan Paul Limited, London.

\bibitem[Brown, 2012]{LockeSolidSouls}
Brown, D.~K. (2012).
\newblock Locke's solid souls.
\newblock {\em Open Journal of Philosophy}, 2(4):228--234.

\bibitem[Brown and Holland, 2004]{BrownHollandNoether1}
Brown, H. and Holland, P. (2004).
\newblock Dynamical versus variational symmetries: {Understanding Noether's}
  first theorem.
\newblock {\em Molecular Physics}, 102:1133--1139.

\bibitem[Bunge, 1980]{Bunge}
Bunge, M. (1980).
\newblock {\em The Mind-Body Problem: A Psychobiological Approach}.
\newblock Pergamon, Oxford.

\bibitem[Burge, 2007]{BurgeMind}
Burge, T. (2007).
\newblock Mind-body causation and explanatory practice.
\newblock In {\em Foundations of Mind: Philosophical Essays, Volume 2}, pages
  344--382. Clarendon Press, Oxford.
\newblock {Reprinted (with new postscript) from John Heil and Alfred R. Mele,
  editors, \emph{Mental Causation}, Clarendon Press, Oxford (1993), pp.
  77-120}.

\bibitem[Butterfield, 1997]{ButterfieldPsychophysics}
Butterfield, J. (1997).
\newblock Quantum curiosities of psychophysics.
\newblock In Cornwell, J., editor, {\em Consciousness and Human Identity},
  pages 122--159. Oxford University Press, Oxford.
\newblock http://philsci-archive.pitt.edu/193/.

\bibitem[Carde{\~{n}}a et~al., 2015]{ParapsychologyHandbook21st}
Carde{\~{n}}a, E., Palmer, J., and Marcusson-Clavertz, D., editors (2015).
\newblock {\em Parapsychology: A Handbook for the 21st Century}.
\newblock McFarland \& Company, Jefferson, North Carolina.

\bibitem[Carroll, 2004]{CarrollSpacetimeGeometry}
Carroll, S. (2004).
\newblock {\em Spacetime and Geometry: An Introduction to General Relativity}.
\newblock Addison-Wesley, San Francisco.

\bibitem[Carroll, 2010]{CarrollEnergyIsNotConserved}
Carroll, S. (2010).
\newblock Energy is not conserved.
\newblock {\em Discover: The Magazine of Science, Technology and the Future}.
\newblock
  http://blogs.discovermagazine.com/cosmicvariance/2010/02/22/energy-is-not-co%
nserved/\#.WaAUO2d3FyA.

\bibitem[Carroll, 2016]{CarrollBigPicture}
Carroll, S. (2016).
\newblock {\em The Big Picture: On the Origins of Life, Meaning, and the
  Universe Itself}.
\newblock Dutton, New York.

\bibitem[Carroll, 2018]{CarrollFacebookReligion}
Carroll, S. (2018).
\newblock Poetic naturalism.
\newblock https://www.preposterousuniverse.com/poetic-naturalism/, accessed 9
  November 2018.

\bibitem[Carroll, 2011]{CarrollMindEnergy}
Carroll, S.~M. (2011).
\newblock Physics and the immortality of the soul.
\newblock {\em Scientific American Guest Blog}.
\newblock
  http://blogs.scientificamerican.com/guest-blog/physics-and-the-immortality-o%
f-the-soul/, May 23, 2011.

\bibitem[Cattani and {De Maria}, 1993]{Cattani}
Cattani, C. and {De Maria}, M. (1993).
\newblock Conservation laws and gravitational waves in {General Relativity}
  (1915-1918).
\newblock In Earman, J., Janssen, M., and Norton, J.~D., editors, {\em The
  Attraction of Gravitation: New Studies in the History of General Relativity},
  volume~5 of {\em Einstein Studies, editors Don Howard and John Stachel},
  pages 63--87. Birkh\"{a}user, Boston.

\bibitem[Chang et~al., 1999]{NesterQuasiPseudo}
Chang, C.-C., Nester, J.~M., and Chen, C.-M. (1999).
\newblock Pseudotensors and quasilocal energy-momentum.
\newblock {\em Physical Review Letters}, 83:1897--1901.
\newblock gr-qc/9809040.

\bibitem[Clarke, 2014]{ClarkeSoul}
Clarke, P. G.~H. (2014).
\newblock Neuroscience, quantum indeterminism and the {Cartesian} soul.
\newblock {\em Brain and Cognition}, 84:109--117.

\bibitem[Collins, 2008]{CollinsEnergy}
Collins, R. (2008).
\newblock Modern physics and the energy-conservation objection to mind-body
  dualism.
\newblock {\em American Philosophical Quarterly}, 45:31--42.

\bibitem[Collins, 2011]{CollinsSoulHypothesis}
Collins, R. (2011).
\newblock The energy of the soul.
\newblock In Baker, M.~C. and Goetz, S., editors, {\em The Soul Hypothesis:
  Investigations into the Existence of the Soul}, pages 123--133. Continuum,
  New York.

\bibitem[Cooperstock, 2000]{CooperstockEnergy}
Cooperstock, F.~I. (2000).
\newblock The role of energy and a new approach to gravitational waves in
  general relativity.
\newblock {\em Annals of Physics}, 282:115--137.
\newblock arXiv:gr-qc/9904046.

\bibitem[Corcoran, 2006]{Corcoran}
Corcoran, K.~J. (2006).
\newblock {\em Rethinking Human Nature: A Christian Materialist Alternative to
  the Soul}.
\newblock Baker Academic, Grand Rapids.

\bibitem[Craig, 1986]{CraigMiraclesGerman}
Craig, W.~L. (1986).
\newblock The problem of miracles: {A} historical and philosophical
  perspective.
\newblock In Wenham, D. and Blomberg, C., editors, {\em The Miracles of Jesus},
  Gospel Perspectives, Volume 6, pages 9--40. JSOT Press, Sheffield.
\newblock http://www.leaderu.com/offices/billcraig/docs/miracles.html.

\bibitem[Crane, 2001]{CraneElements}
Crane, T. (2001).
\newblock {\em Elements of Mind: An Introduction to the Philosophy of Mind}.
\newblock Oxford University Press, Oxford.

\bibitem[Cucu and Pitts, 2019]{EnergyMentalCucuLowe}
Cucu, A.~C. and Pitts, J.~B. (2019).
\newblock How dualists should (not) respond to the objection from energy
  conservation.
\newblock {\em Mind and Matter}, 17:95--121.

\bibitem[Curiel, 2000]{CurielCauseGR}
Curiel, E. (2000).
\newblock The constraints {General Relativity} places on physicalist accounts
  of causality.
\newblock {\em Theoria}, 15:33--58.

\bibitem[Dewar and Weatherall, 2018]{DewarWeatherallGravitationalEnergyNewton}
Dewar, N. and Weatherall, J.~O. (2018).
\newblock On gravitational energy in {Newtonian} theories.
\newblock {\em Foundations of Physics}, 48:558--578.

\bibitem[DeWitt, 2003]{DeWittGlobalQFT}
DeWitt, B. (2003).
\newblock {\em The Global Approach to Quantum Field Theory}.
\newblock Clarendon Press, Oxford.
\newblock Two Volumes.

\bibitem[Dowe, 2000a]{DoweDefended}
Dowe, P. (2000a).
\newblock The conserved quantity theory defended.
\newblock {\em Theoria}, 15:11--31.

\bibitem[Dowe, 2000b]{DowePhysicalCausation}
Dowe, P. (2000b).
\newblock {\em Physical Causation}.
\newblock Cambridge University Press, Cambridge.

\bibitem[Ducasse, 1960]{DucasseDualism}
Ducasse, C. (1960).
\newblock In defense of dualism.
\newblock In Hook, S., editor, {\em Dimensions of Mind: A Symposium}, pages
  85--90. New York University Press, New York.

\bibitem[Duerr, 2019]{DuerrFantasticBeasts}
Duerr, P.~M. (2019).
\newblock Fantastic beasts and where (not) to find them: {Local} gravitational
  energy and energy conservation in general relativity.
\newblock {\em Studies in History and Philosophy of Modern Physics}, 65:1--14.

\bibitem[Duff, 1981]{DuffQFTinCSTinconsistent}
Duff, M.~J. (1981).
\newblock Inconsistency of quantum field theory in curved space-time.
\newblock In Isham, C.~J., Penrose, R., and Sciama, D.~W., editors, {\em
  Quantum Gravity 2: A Second Oxford Symposium}, pages 81--105, Oxford.
  Clarendon.

\bibitem[Earman, 2000]{EarmanHume}
Earman, J. (2000).
\newblock {\em Hume's Abject Failure: The Argument against Miracles}.
\newblock Oxford University Press, Oxford.
\newblock http://pitt.edu/~jearman/Earman2000HumeAbjectFailure.pdf.

\bibitem[Eccles, 1987]{EcclesMindwaves}
Eccles, J. (1987).
\newblock Brain and mind: {Two} or one?
\newblock In Blakemore, C. and Greenfield, S., editors, {\em Mindwaves:
  Thoughts on Intelligence, Identity and Consciousness}, pages 293--304.
  Blackwell, Oxford.

\bibitem[Eccles, 1994]{EcclesSelfBrain}
Eccles, J.~C. (1994).
\newblock {\em How the Self Controls Its Brain}.
\newblock Springer, Berlin.

\bibitem[Eddington, 1939]{EddingtonPhilPhysSci}
Eddington, A.~S. (1939).
\newblock {\em The Philosophy of Physical Science}.
\newblock Cambridge University Press, Cambridge.

\bibitem[Elkin, 2009]{ElkinUnderdeterminedODE}
Elkin, V.~I. (2009).
\newblock Reduction of underdetermined systems of ordinary differential
  equations: I.
\newblock {\em Differential Equations}, 45:1721--1731.
\newblock Russian original in \emph{Differentsial'nye Uravneniya}, pp.
  1687-1697.

\bibitem[Fair, 1979]{FairCauseFlowEnergy}
Fair, D. (1979).
\newblock Causation and the flow of energy.
\newblock {\em Erkenntnis}, 14:219--250.

\bibitem[Feigl, 1958]{FeiglMindBody}
Feigl, H. (1958).
\newblock The ``mental'' and the ``physical''.
\newblock In Feigl, H., Scriven, M., and Maxwell, G., editors, {\em Concepts,
  Theories, and the Mind-Body Problem}, volume~II of {\em Minnesota Studies in
  the Philosophy of Science}, pages 370--497. University of Minnesota Press,
  Minneapolis.

\bibitem[Feynman, 1971]{FeynmanUnpublished}
Feynman, R.~P. (1971).
\newblock {\em Lectures on Gravitation, California Institute of Technology,
  Pasadena, California, 1962-63, Lecture notes by Fernando B. Morinigo and
  William G. Wagner}.
\newblock California Institute of Technology.

\bibitem[Feynman et~al., 1995]{Feynman}
Feynman, R.~P., Morinigo, F.~B., Wagner, W.~G., Hatfield, B., Preskill, J., and
  Thorne, K.~S. (1995).
\newblock {\em Feynman Lectures on Gravitation}.
\newblock Addison-Wesley, Reading, Mass.
\newblock Original by California Institute of Technology, 1963.

\bibitem[Fodor, 1998]{FodorMindBody}
Fodor, J. (1998).
\newblock The mind-body problem.
\newblock In Arnold, N.~S., Benditt, T.~M., and Graham, G., editors, {\em
  Philosophy Then and Now}, pages 63--77. Blackwell, Malden, Massachusetts.
\newblock Reprinted from \emph{Scientific American}, {January 1981}, pp.
  114-123.

\bibitem[Foster, 2001]{FosterSoulBody}
Foster, J. (2001).
\newblock A brief defense of the {Cartesian} view.
\newblock In Corcoran, K., editor, {\em Soul, Body, and Survival: Essays on the
  Metaphysics of Human Persons}, pages 15--29. Cornell University Press,
  Ithaca.

\bibitem[Fran\c{c}ois, 2019]{FrancoisArtificialSubstantialGauge}
Fran\c{c}ois, J. (2019).
\newblock Artificial vs substantial gauge symmetries: a criterion and an
  application to the electroweak model.
\newblock {\em Philosophy of Science}, 86:472--496.
\newblock arXiv:1801.00678v3 [physics.hist-ph].

\bibitem[Fulling, 1989]{Fulling}
Fulling, S.~A. (1989).
\newblock {\em Aspects of Quantum Field Thery in Curved Space-time}.
\newblock Cambridge University Press, Cambridge.

\bibitem[Gabriel, 2017]{GabrielNotaBrain}
Gabriel, M. (2017).
\newblock {\em I Am Not a Brain}.
\newblock Polity Press, Cambridge.
\newblock Translation of \emph{Ich ist nicht Gehirn. Philosophie des Geistes
  f\"{u}r das 21. Jahrhundert}, Ullstein Verlag, Berlin (2015).

\bibitem[Garber, 1983]{GarberLeibnizEnergy}
Garber, D. (1983).
\newblock Mind, body, and the laws of nature in {Descartes} and {Leibniz}.
\newblock {\em Midwest Studies in Philosophy}, 8:105--134.

\bibitem[Gates et~al., 1983]{GatesGrisaruRocekSiegel}
Gates, Jr., S.~J., Grisaru, M.~T., Ro\v{c}ek, M., and Siegel, W. (1983).
\newblock {\em Superspace, or One Thousand and One Lessons in Supersymmetry}.
\newblock Benjamin/Cummings, Reading, Massachusetts.
\newblock hep-th/0108200.

\bibitem[Gibb, 2010]{GibbConservation}
Gibb, S. (2010).
\newblock Closure principles and the laws of conservation of energy and
  momentum.
\newblock {\em Dialectica}, 64:363--384.

\bibitem[Gorelik, 2002]{GorelikConservation}
Gorelik, G. (2002).
\newblock The problem of conservation laws and the {Poincar\'{e}} quasigroup in
  {General Relativity}.
\newblock In Balashov, Y. and Vizgin, V., editors, {\em Einstein Studies in
  Russia}, {Einstein Studies, editors Don Howard and John Stachel}, pages
  17--43. Birkh\"{a}user, Boston.

\bibitem[Graves, 1971]{GravesEnergy}
Graves, J.~C. (1971).
\newblock {\em The Conceptual Foundations of Contemporary Relativity Theory}.
\newblock MIT Press, Cambridge.

\bibitem[Griffin, 1997]{GriffinParapsychology}
Griffin, D.~R. (1997).
\newblock {\em Parapsychology, Philosophy, and Spirituality: A Postmodern
  Exploration}.
\newblock State University of New York Press, Albany.

\bibitem[Grosso, 2016]{LevitationGrossoCopertino}
Grosso, M. (2016).
\newblock {\em The Man Who Could Fly: St. Joseph of Copertino and the Mystery
  of Levitation}.
\newblock Rowman \& Littlefield, Lanham, Maryland.

\bibitem[Hack, 2016]{HackQFT}
Hack, T.-P. (2016).
\newblock {\em Cosmological Applications of Algebraic Quantum Field Theory in
  Curved Spacetimes}.
\newblock Springer, Cham, Switzerland.

\bibitem[Halvorson, 2011]{HalvorsonSoulHypothesis}
Halvorson, H. (2011).
\newblock The measure of all things: {Quantum} mechanics and the soul.
\newblock In Baker, M.~C. and Goetz, S., editors, {\em The Soul Hypothesis:
  Investigations into the Existence of the Soul}, pages 138--163. Continuum,
  New York.

\bibitem[Hamilton, 1834]{HamiltonConservation}
Hamilton, W.~R. (1834).
\newblock On a general method in dynamics; by which the study of the motions of
  all free systems of attracting or repelling points is reduced to the search
  and differentiation of one central relation, or characteristic function.
\newblock {\em Philosophical Transactions of the Royal Society of London},
  124:247--308.

\bibitem[Hasker, 2001]{HaskerCorcoran}
Hasker, W. (2001).
\newblock Persons as emergent substances.
\newblock In Corcoran, K., editor, {\em Soul, Body, and Survival: Essays on the
  Metaphysics of Human Persons}, pages 107--119. Cornell University Press,
  Ithaca.

\bibitem[Hawthorne and Uzquiano, 2011]{HawthorneAngelsNeedle}
Hawthorne, J. and Uzquiano, G. (2011).
\newblock How many angels can dance on the point of a needle? {Transcendental}
  theology meets modal metaphysics.
\newblock {\em Mind}, 120(477):53--81.

\bibitem[Heidelberger, 2003]{FeiglMindBodyLogicalEmpiricism}
Heidelberger, M. (2003).
\newblock The mind-body problem in the origin of {Logical Empiricism}: {Herbert
  Feigl} and psychophysical parallelism.
\newblock In Parrini, P., Salmon, W., and Salmon, M.~H., editors, {\em Logical
  Empiricism: Historical and Contemporary Perspectives}, pages 233--262.
  Pittsburgh University Press, Pittsburgh.

\bibitem[Heidelberger, 2004]{HeidelbergerFechner}
Heidelberger, M. (2004).
\newblock {\em Nature From Within: Gustav Theodor Fechner And His
  Psychophysical Worldview}.
\newblock University of Pittsburgh Press, Pittsburgh.

\bibitem[Hoefer, 2000]{Hoefer}
Hoefer, C. (2000).
\newblock Energy conservation in {GTR}.
\newblock {\em Studies in History and Philosophy of Modern Physics},
  31:187--199.

\bibitem[Hossenfelder, 2016]{HossenfelderEnergy}
Hossenfelder, S. (Wednesday, October 19, 2016).
\newblock {Dear Dr B: Where} does dark energy come from and what's it made of?
\newblock {\em BackReAction}.
\newblock
  http://backreaction.blogspot.com/2016/10/dear-dr-b-where-does-dark-energy-co%
me.html.

\bibitem[Jacobi, 1996]{Jacobi}
Jacobi, C. G.~J. (1996).
\newblock {\em Vorlesungen \"{u}ber analytische Mechanik, Berlin 1847/8}.
\newblock Deutsche Mathematiker-Vereinigung. Vieweg, Braunschweig.
\newblock Edited by {Helmut Pulte}.

\bibitem[Kaku, 1993]{Kaku}
Kaku, M. (1993).
\newblock {\em Quantum Field Theory: A Modern Introduction}.
\newblock Oxford University, New York.

\bibitem[Kastrup, 1987]{KastrupNoetherKleinLie}
Kastrup, H.~A. (1987).
\newblock The contributions of {Emmy Noether, Felix Klein and Sophus Lie} to
  the modern concept of symmetries in physical systems.
\newblock In Doncel, M.~G., Hermann, A., Michel, L., and Pais, A., editors,
  {\em Symmetries in Physics (1600-1980): Proceedings, 1st International
  Meeting on the History of Scientific Ideas, Sant Feliu de Gu\'{i}xols, Spain,
  September 20-26, 1983}, pages 113--163. Seminari d'Hist\`{o}ria de les
  Ci\'{e}ncies, Universitat Aut\'{o}noma de Barcelona, Bellaterra, Barcelona.
\newblock http://www.desy.de/~hkastrup/reprints/.

\bibitem[Kennefick, 2007]{KennefickWaves}
Kennefick, D. (2007).
\newblock {\em Traveling at the Speed of Thought: Einstein and the Quest for
  Gravitational Waves}.
\newblock Princeton University Press, Princeton.

\bibitem[Kent, 2018]{KentQuantumQualia}
Kent, A. (2018).
\newblock Quanta and qualia.
\newblock {\em Foundations of Physics}, 48:1021--1037.
\newblock Contribution to panel discussion at FQXi 2016 International
  Conference. arXiv:1608.04804v4 [quant-ph].

\bibitem[Kim, 2003]{KimLonelySouls}
Kim, J. (2003).
\newblock Lonely souls: {Causality} and substance dualism.
\newblock In O'Connor, T. and Robb, D., editors, {\em Philosophy of Mind:
  Contemporary Readings}. Routledge, London.
\newblock Slightly modified from {Kevin Corcoran}, editor, {\emph{Soul, Body
  and Survival: Essays on the Metaphysics of Human Persons}, Cornell University
  Press, Ithaca}, 2001, pp. 30-43.

\bibitem[Klein, 1971]{KleinIsland}
Klein, O. (1971).
\newblock Arguments concerning relativity and cosmology.
\newblock {\em Science}, 171:339--345.

\bibitem[Kosmann-Schwarzbach, 2011]{KosmannSchwarzbachNoether}
Kosmann-Schwarzbach, Y. (2011).
\newblock {\em The Noether Theorems: Invariance and Conservation Laws in the
  Twentieth Century}.
\newblock Springer, New York.
\newblock {Translated by Bertram E. Schwarzbach}.

\bibitem[Ladyman et~al., 2007]{LadymanRoss}
Ladyman, J., Ross, D., Spurrett, D., and Collier, J. (2007).
\newblock {\em Every Thing Must Go: Metaphysics Naturalized}.
\newblock Oxford University Press, Oxford.

\bibitem[Lagrange, 1811]{LagrangeMecaniqueAnalytiqueNouvelle}
Lagrange, J.-L. (1811).
\newblock {\em M\'{e}canique Analytique}, volume~1.
\newblock Courcier, Paris, revised edition.
\newblock Google Books.

\bibitem[Lagrange, 1997]{LagrangeEnglish}
Lagrange, J.~L. (1997).
\newblock {\em Analytical Mechanics: Translated from the \emph{M\'{e}canique
  analytique, nouvelle \'{e}dition} of 1811}.
\newblock Boston Studies in the Philosophy of Science, Volume 191. Kluwer
  Academic, Dordrecht.
\newblock {Translated and edited by Auguste Boissonnade and Victor N.
  Vagliente}.

\bibitem[Lakatos, 1971]{LakatosHistory}
Lakatos, I. (1971).
\newblock History of science and its rational reconstruction.
\newblock In Buck, R.~C. and Cohen, R.~S., editors, {\em PSA: Proceedings of
  the Biennial Meeting of the Philosophy of Science Association, 1970}, Boston
  Studies in the Philosophy of Science, pages 91--136. D. Reidel, Dordrecht.

\bibitem[Lam, 2010]{LamCauseConservedGR}
Lam, V. (2010).
\newblock Metaphysics of causation and physics of general relativity.
\newblock {\em Humana Mente}, (13):61--80.

\bibitem[Lam, 2011]{LamEnergy}
Lam, V. (2011).
\newblock Gravitational and non-gravitational energy: {The} need for background
  structures.
\newblock {\em Philosophy of Science}, 78:1012--1023.

\bibitem[Landau and Lifshitz, 1975]{Landau}
Landau, L.~D. and Lifshitz, E.~M. (1975).
\newblock {\em The Classical Theory of Fields}.
\newblock Pergamon, Oxford, fourth revised {English} edition.
\newblock Translated by Morton Hamermesh.

\bibitem[Lange, 2002]{LangePhilPhys}
Lange, M. (2002).
\newblock {\em An Introduction to The Philosophy of Physics: Locality, Fields,
  Energy, and Mass}.
\newblock John Wiley \& Sons, Malden, Massachusetts.

\bibitem[Leibniz, 1981]{LeibnizNewEssays}
Leibniz, G.~W. (1981).
\newblock {\em New Essays on Human Understanding}.
\newblock Cambridge University Press, Cambridge.
\newblock Translated and edited by Peter Remnant and Jonathan Bennett.

\bibitem[Leibniz, 1985]{LeibnizTheodicy}
Leibniz, G.~W. (1985).
\newblock {\em Theodicy: Essays on the Goodness of God and the Freedom of Man
  and the Origin of Evil}.
\newblock Open Court, La Salle, Illinois.
\newblock Translator E. M. Huggard.

\bibitem[Leibniz, 1997]{LeibnizNewSystem1695FirstExp}
Leibniz, G.~W. (1997).
\newblock {[First] Explanation of the New System of the Com­munication between
  Substances}, in reply to what was said of it in the \emph{{Journal}} for 12
  {September 1695}.
\newblock In Woolhouse, R.~S. and Francks, R., editors, {\em Leibniz's `New
  System' and Associated Contemporary Texts}, pages 47--52. Clarendon Press,
  Oxford.
\newblock Original April 1696.

\bibitem[Lord, 1976]{LordTensors}
Lord, E.~A. (1976).
\newblock {\em Tensors, Relativity and Cosmology}.
\newblock Tata McGraw-Hill Publishing Co., New Delhi.

\bibitem[Lowe, 1992]{LowePsychophysical}
Lowe, E.~J. (1992).
\newblock The problem of psychophysical causation.
\newblock {\em Australasian Journal of Philosophy}, 70:263--276.

\bibitem[Lowe, 1996]{LoweExperience}
Lowe, E.~J. (1996).
\newblock {\em Subjects of Experience}.
\newblock Cambridge University Press, Cambridge.

\bibitem[Lowe, 2003]{LoweInvisibility}
Lowe, E.~J. (2003).
\newblock Physical causal closure and the invisibility of mental causation.
\newblock In Walter, S. and Heckmann, H.-D., editors, {\em Physicalism and
  Mental Causation: The Metaphysics of Mind and Action}, pages 137--154.
  Imprint Academic, Exeter.

\bibitem[Lycan, 2009]{LycanDualism}
Lycan, W.~G. (2009).
\newblock Giving dualism its due.
\newblock {\em Australasian Journal of Philosophy}, 87:551--563.

\bibitem[Manton and Sutcliffe, 2004]{SolitonsManton}
Manton, N. and Sutcliffe, P. (2004).
\newblock {\em Topological Solitons}.
\newblock Cambridge University Press, Cambridge.

\bibitem[McGinn, 1999]{McGinnMysteriousFlame}
McGinn, C. (1999).
\newblock {\em The Mysterious Flame: Conscious Minds in a Material World}.
\newblock Basic Books, New York.

\bibitem[Merricks, 1999]{MerricksResurrection}
Merricks, T. (1999).
\newblock The resurrection of the body and the life everlasting.
\newblock In Murray, M., editor, {\em Reason for the Hope Within}, pages
  261--286. Eerdmans, Grand Rapids.

\bibitem[Misner et~al., 1973]{MTW}
Misner, C., Thorne, K., and Wheeler, J.~A. (1973).
\newblock {\em Gravitation}.
\newblock Freeman, New York.

\bibitem[Mohrhoff, 1997]{MohrhoffInteractionism}
Mohrhoff, U. (1997).
\newblock Interactionism, energy conservation, and the violation of physical
  laws.
\newblock {\em Physics Essays}, 10:651--665.

\bibitem[Montero, 2006]{MonteroEnergy}
Montero, B. (2006).
\newblock What does the conservation of energy have to do with physicalism?
\newblock {\em Dialectica}, 60:383--396.

\bibitem[Moreland and Craig, 2003]{MorelandCraigWorldview}
Moreland, J.~P. and Craig, W.~L. (2003).
\newblock {\em Philosophical Foundations for a Christian Worldview}.
\newblock InterVarsity, Downers Grove, Illinois.

\bibitem[Moreland and Rae, 2000]{MorelandRaeBodySoul}
Moreland, J.~P. and Rae, S.~B. (2000).
\newblock {\em Body \& Soul: Human Nature \& the Crisis in Ethics}.
\newblock Intervarsity Press, Downers Grove, Illinois.

\bibitem[Motl, 2010]{MotlEnergyNotConserved}
Motl, L. (August 11, 2010).
\newblock Why and how energy is not conserved in cosmology.
\newblock {\em The Reference Frame}.
\newblock
  https://motls.blogspot.com/2010/08/why-and-how-energy-is-not-conserved-in.ht%
ml?m=1.

\bibitem[Murphy, 1998]{MurphySoul}
Murphy, N. (1998).
\newblock Human nature: {Historical}, scientific, and religious issues.
\newblock In Brown, W.~S., Murphy, N., and Maloney, H.~N., editors, {\em
  Whatever Happened to the Soul? Scientific and Theological Portraits of Human
  Nature}, pages 1--30. Fortress Press, Minneapolis.

\bibitem[Nathan, 1991]{McTaggartImmaterialismNathan}
Nathan, N. M.~L. (1991).
\newblock {McTaggart's} immaterialism.
\newblock {\em The Philosophical Quarterly}, 41:442--456.

\bibitem[Nester, 2004]{NesterQuasi}
Nester, J.~M. (2004).
\newblock General pseudotensors and quasilocal quantities.
\newblock {\em Classical and Quantum Gravity}, 21:S261--S280.

\bibitem[Nijenhuis, 1952]{Nijenhuis}
Nijenhuis, A. (1952).
\newblock {\em Theory of the Geometric Object}.
\newblock PhD thesis, University of Amsterdam.

\bibitem[Noether, 1918]{Noether}
Noether, E. (1918).
\newblock {Invariante Variationsprobleme}.
\newblock {\em Nachrichten der K\"{o}niglichen Gesellschaft der Wissenschaften
  zu G\"{o}ttingen, Mathematisch-Physikalische Klasse}, pages 235--257.
\newblock Translated as ``{Invariant} Variation Problems'' by {M. A. Tavel,
  \emph{Transport Theory and Statistical Physics}} {\bf 1} pp. 183-207 (1971),
  {LaTeXed by Frank Y. Wang}, arXiv:physics/0503066 [physics.hist-ph].

\bibitem[Ogievetski\u{i} and Polubarinov, 1965]{OPspinor}
Ogievetski\u{i}, V.~I. and Polubarinov, I.~V. (1965).
\newblock Spinors in gravitation theory.
\newblock {\em Soviet Physics JETP}, 21:1093--1100.

\bibitem[Ogievetsky and Polubarinov, 1965]{OP}
Ogievetsky, V.~I. and Polubarinov, I.~V. (1965).
\newblock Interacting field of spin 2 and the {Einstein} equations.
\newblock {\em Annals of Physics}, 35:167--208.

\bibitem[Parker and Toms, 2009]{ParkerQFTiCST}
Parker, L.~E. and Toms, D.~J. (2009).
\newblock {\em Quantum Field Theory in Curved Spacetime: Quantized Fields and
  Gravity}.
\newblock Cambridge University Press, Cambridge.

\bibitem[Pasnau, 2011]{Pasnau}
Pasnau, R. (2011).
\newblock {\em Metaphysical Themes 1274-1671}.
\newblock Clarendon Press, Oxford.

\bibitem[Pasricha, 1990]{PasrichaEmpirical}
Pasricha, S. (1990).
\newblock {\em Claims of Reincarnation: An Empirical Study of Cases in India}.
\newblock Harman Publishing House, New Delhi.

\bibitem[Pasricha, 2008]{PasrichaEvidence}
Pasricha, S.~K. (2008).
\newblock {\em Can the Mind Survive beyond Death? In Pursuit of Scientific
  Evidence. Volume I: Reincarnation Research}.
\newblock Harman Publishing House, New Delhi.

\bibitem[Penrose, 1994]{PenroseShadows}
Penrose, R. (1994).
\newblock {\em Shadows of the Mind: A Search for the Missing Science of
  Consciousness}.
\newblock Oxford University Press, Oxford.

\bibitem[Peskin and Schroeder, 1995]{PeskinSchroeder}
Peskin, M.~E. and Schroeder, D.~V. (1995).
\newblock {\em An Introduction to Quantum Field Theory}.
\newblock Addison-Wesley, Reading, Massachusetts.

\bibitem[Pitts, 2008]{PittsKalam}
Pitts, J.~B. (2008).
\newblock Why the big bang singularity does not help the kal\={a}m cosmological
  argument for theism.
\newblock {\em The British Journal for the Philosophy of Science}, 59:675--708.
\newblock Reprinted in {Paul Copan with William Lane Craig, editors, \emph{The
  Kal\={a}m Cosmological Argument. Volume 2: Scientific Evidence for the
  Beginning of the Universe}. Bloomsbury Academic, New York (2018), pp.
  80-109.}

\bibitem[Pitts, 2009a]{PittsArtificial}
Pitts, J.~B. (2009a).
\newblock Empirical equivalence, artificial gauge freedom and a generalized
  {Kretschmann} objection.
\newblock http://philsci-archive.pitt.edu/archive/00004995/; arXiv:0911.5400.

\bibitem[Pitts, 2009b]{EnergyGravityConf}
Pitts, J.~B. (2009b).
\newblock Gauge-invariant localization of infinitely many gravitational
  energies from all possible auxiliary structures, or, why pseudotensors are
  okay.
\newblock {Proceedings of the DPF-2009 Conference, Division of Particles and
  Fields, American Physical Society, Wayne State University, Detroit, Michigan;
  arXiv:0910.3320 [hep-th]}.

\bibitem[Pitts, 2010]{EnergyGravity}
Pitts, J.~B. (2010).
\newblock Gauge-invariant localization of infinitely many gravitational
  energies from all possible auxiliary structures.
\newblock {\em General Relativity and Gravitation}, 42:601--622.
\newblock 0902.1288 [gr-qc].

\bibitem[Pitts, 2012]{PittsSpinor}
Pitts, J.~B. (2012).
\newblock The nontriviality of trivial general covariance: {How}  electrons
  restrict `time' coordinates, spinors (almost) fit into tensor calculus, and
  $7/16$ of a tetrad is surplus structure.
\newblock {\em Studies in History and Philosophy of Modern Physics}, 43:1--24.
\newblock arXiv:1111.4586.

\bibitem[Pitts, 2013]{TAM2013TimeandFermionsConfProc}
Pitts, J.~B. (2013).
\newblock Time and fermions: {General} covariance \emph{vs.} {Ockham's} razor
  for spinors.
\newblock In O'Loughlin, M., Stani\v{c}, S., and Veberi\v{c}, D., editors, {\em
  Proceedings of the 4th International Conference on Time and Matter, 4-8 March
  2013, Venice, Italy}, pages 185--198. University of Nova Gorica Press, Nova
  Gorica, Slovenia.

\bibitem[Pitts, 2016a]{ConverseHilbertian}
Pitts, J.~B. (2016a).
\newblock Einstein's equations for spin $2$ mass $0$ from {Noether's} converse
  {Hilbertian} assertion.
\newblock {\em Studies in History and Philosophy of Modern Physics}, 56:60--69.
\newblock PhilSci; arxiv:1611.02673 [physics.hist-ph].

\bibitem[Pitts, 2016b]{EinsteinEnergyStability}
Pitts, J.~B. (2016b).
\newblock Einstein's physical strategy, energy conservation, symmetries, and
  stability: {``but Grossmann \& I believed that the conservation laws were not
  satisfied''}.
\newblock {\em Studies in History and Philosophy of Modern Physics}, 54:52--72.
\newblock PhilSci; arxiv.org/1604.03038 [physics.hist-ph].

\bibitem[Pitts, 2016c]{MassiveGravity3}
Pitts, J.~B. (2016c).
\newblock Universally coupled massive gravity, {III}: {dRGT-Maheshwari} pure
  spin-$2$, {Ogievetsky-Polubarinov} and arbitrary mass terms.
\newblock {\em Annals of Physics}, 365:73--90.
\newblock arXiv:1505.03492 [gr-qc].

\bibitem[Pitts, 2019a]{EnergyMental}
Pitts, J.~B. (2019a).
\newblock Conservation laws and the philosophy of mind: Opening the black box,
  finding a mirror.
\newblock {\em Philosophia}.
\newblock doi:10.1007/s11406-019-00102-7.

\bibitem[Pitts, 2019b]{EnergyMentalHistory}
Pitts, J.~B. (2019b).
\newblock The mind-body problem and conservation laws: The growth of physical
  understanding?
\newblock Under review.

\bibitem[Plantinga, 2007]{PlantingaMaterialism}
Plantinga, A. (2007).
\newblock Materialism and {Christian} belief.
\newblock In van Inwagen, P. and Zimmerman, D., editors, {\em Persons: Human
  and Divine}, pages 99--141. Oxford University Press, New York.

\bibitem[Pons, 2010]{PonsSubstituting}
Pons, J.~M. (2010).
\newblock Substituting fields within the action: {Consistency} issues and some
  applications.
\newblock {\em Journal of Mathematical Physics}, 51:122903.
\newblock arXiv:0909.4151.

\bibitem[Pons, 2011]{PonsEnergy}
Pons, J.~M. (2011).
\newblock Noether symmetries, energy-momentum tensors and conformal invariance
  in classical field theory.
\newblock {\em Journal of Mathematical Physics}, 52(1):012904.
\newblock arXiv:0902.4871.

\bibitem[Popper and Eccles, 1983]{PopperEccles}
Popper, K. and Eccles, J.~C. (1983).
\newblock {\em The Self and Its Brain: An Argument for Interactionism}.
\newblock Routledge, second edition.

\bibitem[Read, 2018]{ReadEnergy}
Read, J. (2018).
\newblock Functional gravitational energy.
\newblock {\em The British Journal for the Philosophy of Science}.
\newblock https://doi.org/10.1093/bjps/axx048.

\bibitem[Reid, 2008]{ReidSpatialSpirits}
Reid, J.~W. (2008).
\newblock The spatial presence of spirits among the {Cartesians}.
\newblock {\em Journal of the History of Philosophy}, 46:91--118.

\bibitem[Rescher, 1991]{LeibnizMonadMachine}
Rescher, N., editor (1991).
\newblock {\em G. W. Leibniz's Monadology: An Edition for Students}.
\newblock University of Pittsburgh Press, Pittsburgh.

\bibitem[{Romero-Maltrana}, 2015]{ConservationSymmetryRomero}
{Romero-Maltrana}, D. (2015).
\newblock Symmetries as by-products of conserved quantities.
\newblock {\em Studies in History and Philosophy of Modern Physics},
  52:358--368.

\bibitem[Rovelli, 2004]{RovelliBook}
Rovelli, C. (2004).
\newblock {\em Quantum Gravity}.
\newblock Cambridge University Press, Cambridge.

\bibitem[Rueger, 1998]{RuegerCauseEnergy}
Rueger, A. (1998).
\newblock Local theories of causation and the a posteriori identification of
  the causal relation.
\newblock {\em Erkenntnis}, 48:25--38.

\bibitem[Ruse, 2018]{RuseLeibnizMachineReligionNotNewAtheism}
Ruse, M. (2018).
\newblock {I'm an Atheist. But thank God I'm not a New Atheist}.
\newblock {\em Premier Christianity}.
\newblock 25 September;
  https://www.premierchristianity.com/Blog/I-m-an-atheist.-But-thank-God-I-m-n%
ot-a-New-Atheist.

\bibitem[Russell, 1927]{RussellMatter}
Russell, B. (1927).
\newblock {\em The Analysis of Matter}.
\newblock Kegan Paul, Trench, Trubner \& Co., London; Harcourt, Brace and
  Company, New York.

\bibitem[Sagan, 1979]{SaganBrocaStevenson}
Sagan, C. (1979).
\newblock {\em Broca's Brain: The Romance of Science}.
\newblock Hodder and Stoughton, London.

\bibitem[Schmaltz, 2008]{SchmaltzDescartesCausation}
Schmaltz, T.~M. (2008).
\newblock {\em Descartes on Causation}.
\newblock Oxford University Press, New York.

\bibitem[Schouten, 1954]{Schouten}
Schouten, J.~A. (1954).
\newblock {\em Ricci-Calculus: An Introduction to Tensor Analysis and Its
  Geometrical Applications}.
\newblock Springer, Berlin, second edition.
\newblock http://link.springer.com/book/10.1007

\bibitem[Schr\"{o}dinger, 1918]{SchrodingerEnergy}
Schr\"{o}dinger, E. (1918).
\newblock Die {Energiekomponenten des Gravitationsfeldes}.
\newblock {\em Physikalische Zeitschrift}, 19:4--7.

\bibitem[Searle, 2004]{SearleMind}
Searle, J.~R. (2004).
\newblock {\em Mind: A Brief Introduction}.
\newblock Oxford University Press, Oxford.

\bibitem[Siegel, 2018]{SiegelEnergy}
Siegel, E. (March 10, 2018).
\newblock Ask {Ethan: Where} is the line between mathematics and physics?
\newblock {\em Forbes: Starts with a Bang}.

\bibitem[Smith, 2001]{CartesianEccles}
Smith, C. U.~M. (2001).
\newblock Renatus renatus: {The Cartesian} tradition in {British} neuroscience
  and the neurophilosophy of {John Carew Eccles}.
\newblock {\em Brain and Cognition}, 46:364--372.

\bibitem[Smoller and Temple, 2003]{SmollerTemple}
Smoller, J. and Temple, B. (2003).
\newblock Shock-wave cosmology inside a black hole.
\newblock {\em Proceedings of the National Academy of Sciences of the United
  States of America}, 100:11216--11218.

\bibitem[Smythies and Beloff, 1989]{SmythiesBeloff}
Smythies, J.~R. and Beloff, J., editors (1989).
\newblock {\em The Case for Dualism}.
\newblock University Press of Virginia, Charlottesville.

\bibitem[Sorkin, 1977]{SorkinStress}
Sorkin, R. (1977).
\newblock On stress-energy tensors.
\newblock {\em General Relativity and Gravitation}, 8:437--449.

\bibitem[Steinkamp, 2002]{SteinkampBeloff}
Steinkamp, F., editor (2002).
\newblock {\em Parapsychology, Philosophy and the Mind: Essays Honoring John
  Beloff}.
\newblock McFarland \& Co., Jefferson, North Carolina.

\bibitem[Stephani, 1990]{Stephani}
Stephani, H. (1990).
\newblock {\em General Relativity}.
\newblock Cambridge University Press, Cambridge, second edition.

\bibitem[Stevenson, 1983]{CasesReincarnationType}
Stevenson, I. (1975-1983).
\newblock {\em Cases of the Reincarnation Type}, volume v. 1. Ten cases in
  India (1975).--v. 2. Ten cases in Sri Lanka (1977).--v. 3. Twelve cases in
  Lebanon and Turkey (1980).--v. 4. Twelve cases in Thailand and Burma (1983).
\newblock University Press of Virginia, Charlottesville.

\bibitem[Stevenson, 1997]{StevensonReincarnationBiologyBirth}
Stevenson, I. (1997).
\newblock {\em Reincarnation and Biology: A Contribution to the Etiology of
  Birthmarks and Birth Defects. Volume 1: Birthmarks, Volume 2: Birth Defects
  and Other Anomalies}.
\newblock Praeger, Westport, Connecticut.

\bibitem[Szabados, 1991]{SzabadosSparlingHAS}
Szabados, L.~B. (1991).
\newblock Canonical pseudotensors, {Sparling's} form and {Noether} currents.
\newblock http://www.rmki.kfki.hu/~lbszab/doc/sparl11.pdf.

\bibitem[Taliaferro, 1994]{TaliaferroConsciousnessGod}
Taliaferro, C. (1994).
\newblock {\em Consciousness and the Mind of God}.
\newblock Cambridge University Press, Cambridge.

\bibitem[Thompson, 2008]{ThompsonMind}
Thompson, I.~J. (2008).
\newblock Discrete degrees within and between nature and mind.
\newblock In Antonietti, A., Corradini, A., and Lowe, J., editors, {\em
  Psycho-Physical Dualism Today: An Interdisciplinary Approach}, pages 99--123.
  Lexington Books, Lanham, Maryland.

\bibitem[Trautman, 1966]{TrautmanUspekhi}
Trautman, A. (1966).
\newblock The general theory of relativity.
\newblock {\em Soviet Physics Uspekhi}, 89:319--339.

\bibitem[Turing, 1950]{TuringESP}
Turing, A. (1950).
\newblock Computing machinery and intelligence.
\newblock {\em Mind}, 59:433--460.

\bibitem[Vailati, 1993]{ClarkeExtendedSoul}
Vailati, E. (1993).
\newblock Clarke's extended soul.
\newblock {\em Journal of the History of Philosophy}, 31:387--403.

\bibitem[{van Inwagen}, 1995]{PVIDualismMaterialism}
{van Inwagen}, P. (1995).
\newblock Dualism and materialism: {Athens and Jerusalem}?
\newblock {\em Faith and Philosophy}, 12:475--488.

\bibitem[{van Nieuwenhuizen}, 1981]{vanNGauge}
{van Nieuwenhuizen}, P. (1981).
\newblock Classical gauge fixing in quantum field theory.
\newblock {\em Physical Review D}, 24:3315--3318.

\bibitem[{von Wachter}, 2006]{WachterCausalClosureImmaterialThings}
{von Wachter}, D. (2006).
\newblock Why the argument from causal closure against the existence of
  immaterial things is bad.
\newblock In Koskinen, H.~J., Vilkko, R., and Philstr\"{o}m, S., editors, {\em
  Science --- A Challenge to Philosophy?}, pages 113--124. Peter Lang,
  Frankfurt.
\newblock https://epub.ub.uni-muenchen.de/1952/.

\bibitem[Wald, 1994]{WaldQFTiCST}
Wald, R.~M. (1994).
\newblock {\em Quantum Field Theory in Curved Spacetime and Black Hole
  Thermodynamics}.
\newblock University of Chicago Press, Chicago.

\bibitem[Watkins, 1995]{WatkinsPhysicalInfluxCrusius}
Watkins, E. (1995).
\newblock The development of physical influx in early eighteenth-century
  {Germany}: {Gottsched, Knutzen, and Crusius}.
\newblock {\em Review of Metaphysics}, 49:295--339.

\bibitem[Watkins, 1998]{WatkinsHarmonyLeibnizPhysicalInflux}
Watkins, E. (1998).
\newblock From pre-established harmony to physical influx: {Leibniz's}
  reception in eighteenth century {Germany}.
\newblock {\em Perspectives on Science}, 6:136--203.

\bibitem[Westphal, 2016]{Westphal}
Westphal, J. (2016).
\newblock {\em The Mind-Body Problem}.
\newblock MIT Press, Cambridge, Massachusetts.

\bibitem[Weyl, 1922]{WeylSTM}
Weyl, H. (1922).
\newblock {\em Space-Time-Matter}.
\newblock Methuen \& Company, London.
\newblock Translated by {Henry L. Brose} from 4th edition of
  \emph{{Raum-Zeit-Materie}}; reprinted by {Dover, New York} (1952).

\bibitem[Wiebe, 2004]{WiebeGodSpirits}
Wiebe, P.~H. (2004).
\newblock {\em God and Other Spirits: Intimations of Transcendence in Christian
  Experience}.
\newblock Oxford University Press, New York.

\bibitem[Wigner, 1962]{WignerMindBody}
Wigner, E.~P. (1962).
\newblock Remarks on the mind-body problem.
\newblock In Good, I.~J., Mayne, A.~J., and {Maynard Smith}, J., editors, {\em
  The Scientist Speculates: An Anthology of Partly-Baked Ideas}, pages
  284--302. Heinemann, London.

\bibitem[Wong, 2007]{CartesianPsychophysics}
Wong, H.~Y. (2007).
\newblock Cartesian psychophysics.
\newblock In van Inwagen, P. and Zimmerman, D., editors, {\em Persons: Human
  and Divine}, pages 169--195. Oxford University Press, New York.

\bibitem[Woodard, 1984]{WoodardSymmetricTetrad}
Woodard, R.~P. (1984).
\newblock The vierbein is irrelevant in perturbation theory.
\newblock {\em Physics Letters B}, 148:440--444.

\bibitem[Young and Goulet, 1994]{AnthropologyBeingChanged}
Young, D.~E. and Goulet, J.-G., editors (1994).
\newblock {\em Being Changed by Cross-Cultural Encounters: The Anthropology of
  Extraordinary Experience}.
\newblock Broadview Press, Peterborough, Ontario.
\newblock Reprinted by University of Toronto Press, Higher Education Division,
  North York, 2012.

\bibitem[Zimmerman, 2007]{ZimmermanDualism}
Zimmerman, D. (2007).
\newblock Dualism in the philosophy of mind.
\newblock In Borchert, D.~M., editor, {\em Encyclopedia of Philosophy}, pages
  113--122. Macmillan, New York, second edition.
\newblock http://fas-philosophy.rutgers.edu/zimmerman/Dualism.in.Mind.pdf.

\end{thebibliography}
\end{document}